\documentclass[journal]{IEEEtran}
\usepackage{amsmath,graphicx}
\usepackage{graphics}
\usepackage{algorithm,algorithmicx,algpseudocode}
\usepackage{amsfonts}
\usepackage{amssymb}
\usepackage{epstopdf}
\usepackage[cal=cm]{mathalfa}
\usepackage[figuresright]{rotating}
\usepackage{subfigure,url}
\usepackage{tabto}

\usepackage{jmei_custom}
\graphicspath{{new_figs/},{revise_resubmit_figs/},{revise_resubmit_figs/AQ/}}

\allowdisplaybreaks

\title{Signal Processing on Graphs: Causal Modeling of \textit{Un}structured Data}
\author{Jonathan Mei and Jos\'e M.~F.~Moura
\thanks{The authors are with the Department of Electrical and Computer Engineering, Carnegie Mellon University, Pittsburgh, PA 15213, USA. ph: (412)268-6341; fax: (412)268-3890; e-mails: [jmei1,moura]@ece.cmu.edu.}
\thanks{This work was partially funded by NSF grants CCF 1011903 and CCF 1513936.}
\thanks{This paper appears in: IEEE Transactions on Signal Processing, vol. 65, no. 8, pp. 2077-2092, April 15, 2017.}
\thanks{Print ISSN: 1053-587X; Online ISSN: 1941-0476}
\thanks{DOI: 10.1109/TSP.2016.2634543}}

\begin{document}
%

\maketitle
\begin{abstract}
Many applications collect a large number of time series, for example, the financial data of companies quoted in a stock exchange, the health care data of all patients that visit the emergency room of a hospital, or the temperature sequences continuously measured by weather stations across the US. These data are often referred to as \textit{un}structured. A first task in its analytics is to derive a low dimensional representation, a graph or discrete manifold, that describes well the \textit{inter}relations among the time series and their \textit{intra}relations across time. This paper presents a computationally tractable algorithm for estimating this graph that structures the data. The resulting graph is directed and weighted, possibly capturing causal relations, not just reciprocal correlations as in many existing approaches in the literature. A convergence analysis is carried out. The algorithm is demonstrated on random graph datasets and real network time series datasets, and its performance is compared to that of related methods. The adjacency matrices estimated with the new method are close to the true graph in the simulated data and consistent with prior physical knowledge in the real dataset tested.
\end{abstract}
%
\hspace{.43cm}\textbf{Keywords:}
Graph Signal Processing, Graph Structure, Adjacency Matrix, Network, Time Series, Big Data, Causal
%
\section{Introduction}
\label{sec:intro}
There is an explosion of data, generated, measured, and stored at very fast rates in many disciplines, from finance and social media to geology and biology. Much of this data takes the form of simultaneous, long running time series. Examples include protein-to-protein interactions in organisms, patient records in health care, customer consumptions in (power, water, natural gas) utility companies, cell phone usage for wireless service providers, companies' financial data, and social interactions among individuals in a population. The internet-of-things (IoT) is an imminent source of ever increasing large collections of time series. This data is often referred to as \textit{un}structured.

Networks or graphs are becoming prevalent as models to describe the relationships among the nodes and the data they produce. These low-dimensional graph data representations are used for further analytics, for example, to compute statistics, make inferences, perform signal processing tasks~\cite{jackson_social_2010,murphy_loopy_1999,sandryhaila_discrete_2013}, or quantify how topology influences diffusion in networks of nodes~\cite{newman_networks:_2010}. These methods all use the structure of the \emph{known} graphs to extract knowledge and meaning from the observed data supported on the graphs.

However, in many problems the graph structure itself may be \emph{unknown}, and a first issue is to infer from the data the unknown relations between the nodes. The diversity and \textit{un}structured nature of the data challenges our ability to derive models from first principles; alternatively, because data is abundant, it is of great significance to develop methodologies that, in collaboration with domain experts, assist extracting low-dimensional representations that structure the data. Early work in estimating low-dimensional graph-like structure includes dimensionality reduction approaches such as \cite{roweis_nonlinear_2000,tenenbaum_global_2000}. These methods work on a static snapshot of data and do not incorporate the notion of time.

This paper focuses on estimating the network structure capturing the dependencies among time series in the form of a possibly directed, weighted adjacency matrix~$\A$. Current work on estimating network structure largely associates it with assuming that the process supported by the graph is Markov along the edges~\cite{meinshausen_high-dimensional_2006,ravikumar_high-dimensional_2011} or aims to recover causality~\cite{bolstad_causal_2011} in the sense popularized by Granger~\cite{granger_investigating_1969}.
Our work instead associates the graph with causal network effects, drawing inspiration from the Discrete Signal Processing on Graphs (\DSPG)~framework~\cite{sandryhaila_discrete_2013,sandryhaila_discrete_2014}.

We provide a brief overview of the concepts and notations underlying \DSPG~and then introduce related prior work in section~\ref{sec:Prior} and our new network process in section~\ref{sec:CGP}. Next, we present algorithms to infer the network structure from data generated by such processes in section~\ref{sec:Estim}. We provide analysis on the convergence of our algorithms and the performance of the models for prediction in section~\ref{sec:Conv}. Finally, we show simulation results in section~\ref{sec:Exp} and conclude the paper in section~\ref{sec:Concl}.

\section{Relation to Prior Work}
\label{sec:Prior}

We briefly review \DSPG~ and then describe previous models and methods used to estimate graph structure. Section~\ref{subsec:GraphicalModel} considers sparse Gaussian graphical model selection, Section~\ref{subsec:SpVar} sparse vector autoregression, and Section~\ref{subsec:GSP_Laplacian} other graph signal processing approaches.

\noindent{\textbf{\DSPG~review}}

Consider a graph $G=(\V,\A)$ where the vertex set $\V=\{v_0,\ldots,v_{N-1}\}$ and $\A$ is the weighted adjacency matrix of the graph. A graph signal is defined as a map, $\x : \V \rightarrow \mathbb{C}$, $v_n \mapsto x_n$, that we can write as $\x = \begin{array}{cccc}( x_0 \!&\! x_1 \!&\! \ldots \!&\! x_{N-1} )^T \end{array} \in \mathbb{C}^N$.

We adopt the adjacency matrix $\A$ as a shift~\cite{sandryhaila_discrete_2014}, and assuming shift invariance and that the minimal polynomial $m_\A(x)$ of $\A$ equals its characteristic polynomial, graph filters in \DSPG~are matrix polynomials of the form $h(\A)=h_0 \I + h_1 \A + \ldots + h_{L_h} \A^{L_h}$, where $L_h$ is the polynomial order.

\subsection{Sparse Graphical Model Selection}
\label{subsec:GraphicalModel}

Sparse inverse covariance estimation~\cite{friedman_sparse_2008,chandrasekaran_latent_2012,ravikumar_high-dimensional_2011} combines the Markov property with the assumptions of Gaussianity and symmetry using Graphical Lasso~\cite{friedman_sparse_2008}. Let the data matrix representing all the $K$ observations is given,
\begin{equation}
\label{eq:DataMatrix}
\X=\begin{array}{cccc}\big(\x[0]&\x[1]&\ldots&\x[K-1]\big)\end{array}\in \mathbb{R}^{N\times K},
\end{equation}
where $\x[i] \sim \mathcal{N}(\0,\boldsymbol{\Sigma})$, $\{\x[i]\}$ are i.i.d., and an estimate for $\boldsymbol{\Theta}=\boldsymbol{\Sigma}^{-1}$ is desired. The regularized likelihood function is maximized, leading to

\begin{equation}
\label{eq:GLasso}
\widehat{\boldsymbol{\Theta}}=\underset{\boldsymbol{\Theta}}{\arg\!\min}\; \textrm{tr}\left(\S\boldsymbol{\Theta}\right)-\log\left|\boldsymbol{\Theta}\right|+\lambda\left\|\boldsymbol{\Theta}\right\|_1,
\end{equation}
where $\S=\frac{1}{K}\X\X^T$ is the sample covariance matrix and $\|\boldsymbol{\Theta}\|_1=\sum\limits_{i,j}|\boldsymbol{\Theta}_{ij}|$.

Given time series data generated from a sparse graph process, the inverse covariance matrix $\boldsymbol{\Theta}$ can actually reflect higher order effects and be significantly less sparse than the graph underlying the process. For example, if a process is described by the dynamic equation with sparse state evolution matrix $\A$ and $\|\A\| \le 1$,
\begin{equation*}
\x[k]=\A\x[k-1]+\w[k],
\end{equation*}
where $\w[i]\sim\Ncal(\0,\boldsymbol{\Sigma}_\w)$ is a random noise process that is generated independently from $\w[j]$ for all $i\ne j$, then
\begin{equation*}
\begin{aligned}
\boldsymbol{\Sigma} &=\mathbb{E}\left[\x[k]\x[k]^T\right]=\sum\limits_{i=0}^\infty \A^i \boldsymbol{\Sigma}_\w \left(\A^T\right)^i \\
\Rightarrow \boldsymbol{\Theta}&=\left(\sum\limits_{i=0}^\infty \A^i \boldsymbol{\Sigma}_\w \left(\A^T\right)^i \right)^{-1}.
\end{aligned}
\end{equation*}
Even though the process can be described by sparse matrix $\A$, the true value of $\boldsymbol{\Theta}$ represents powers of $\A$ and need not be as sparse as $\A$. In addition, $\A$ may not be symmetric, i.e., the underlying graph may be directed, while $\boldsymbol{\Theta}$ is symmetric, and the corresponding graph is undirected.

\subsection{Sparse Vector Autoregressive Estimation}
\label{subsec:SpVar}

For time series data, instead of estimating the inverse covariance structure, sparse vector autoregressive (SVAR) estimation~\cite{bach_learning_2004,songsiri_topology_2010,bolstad_causal_2011} recovers matrix coefficients for multivariate processes.
This SVAR model assumes the time series at each node are conditionally independent from each other according to a Markov Random Field (MRF) with adjacency structure given by $\A'\in\{0,1\}^{N\times N}$. This $\A'$ is related to Granger causality~\cite{granger_causality_1988,granger_recent_1988} as shown in~\cite{bolstad_causal_2011}.
This problem assumes given a data matrix of the form in equation~(\ref{eq:DataMatrix}) that is generated by the dynamic equation with sparse evolution matrices $\{\A^{(i)}\}$,\par
{\centering
$\x[k]=\sum\limits_{i=1}^M \A^{(i)}\x[k-i]+\w[k]$,
\par}
\noindent where $\w[i]$ is a random noise process independent from $\w[j]$, $i\ne j$, and $\A^{(i)}$ all have the same sparse structure, $\A'_{ij}=0\Rightarrow\A^{(k)}_{ij}=0$ for all $k$. Then SVAR solves,
\begin{equation}
\begin{aligned}
\{\widehat{\A}^{(i)}\}=\underset{\{\A^{(i)}\}}{\arg\!\min}\;\frac{1}{2}&\sum\limits_{k=M}^{K-1}\left\|\x[k]-\sum\limits_{i=1}^{M} \A^{(i)} \x[k-i] \right\|_2^2\\
+&\lambda\sum\limits_{i,j}\|\a_{ij}\|_2,
\end{aligned}
\label{eq:SVAR}
\end{equation}
where $\a_{ij}=\left(\!\begin{array}{ccc}A^{(1)}_{ij}&\ldots&A^{(M)}_{ij}\end{array}\!\right)^T$ and $\|\a_{ij}\|_2$ promotes sparsity of the $(i,j)$-th entry of each $\A^{(i)}$ simultaneously and thus also sparsity of $\A'$. This optimization can be solved using Group Lasso~\cite{friedman_sparse_2008}.
SVAR estimates multiple weighted graphs $\A^{(i)}$ that can be used to estimate the sparsity structure $\A'$.

In contrast, our model is defined by a single weighted $\A$, a potentially more parsimonious model than~\eqref{eq:SVAR}. The corresponding time filter coefficients (introduced in section~\ref{sec:CGP}) are modeled as graph filters.
Using this single adjacency matrix $\A$ and graph filters to describe the process enables principled analysis using the toolbox provided by the \DSPG~framework.

\subsection{Graph Signal Processing using the Laplacian}
\label{subsec:GSP_Laplacian}

Frameworks for signal processing on graphs using the weighted (and/or possibly normalized) Laplacian matrix rather than the weighted adjacency matrix have been proposed~\cite{shuman_emerging_2013, hammond_wavelets_2011, coifman_diffusion_2006}. For example, with a known graph Laplacian matrix, the polynomial coefficients for optimal graph filters can be learned and used to describe and compress signals~\cite{thanou_learning_2014-1}.

The symmetric Laplacian is positive semidefinite with all real nonnegative eigenvalues and a real orthonormal eigenvector matrix. Most Laplacian-based Graph Signal Processing methods assume the Laplacian to be symmetric and implicitly take advantage of these properties in various ways.

Recently, methods for estimating symmetric Laplacians have been proposed and came to our attention after the submission of this paper~\cite{dong_learning_2014-1,kalofolias_how_2016,pavez_generalized_2016}. These methods estimate Laplacians for independent graph signals with an interpretation similar to the (inverse) covariance matrix in section~\ref{subsec:GraphicalModel}, and do not take into account the temporal structure and dependencies of the data. In addition, these methods all depend implicitly on the symmetry of the Laplacian, which yields an orthonormal eigenbasis. Furthermore, they depend on the conic geometry of the space of symmetric positive semidefinite matrices, which allows the utilization of convex optimization.

Symmetric Laplacians correspond to undirected graphs, having nonnegative real eigenvalues. This may be restrictive in applications, since it assumes symmetric relations among time series data. Asymmetric Laplacians are also now being studied, but they are restricted to having zero row (or column) sums, which is often undesirable.

In contrast, the directed adjacency matrix $\A$ that we assume may have positive as well as negative weights on edges and can have complex eigenvalues. We add that knowing the adjacency matrix does allow us to compute the Laplacian. In this paper, we adopt the adjacency matrix as the basic building block.

\section{Causal Graph Processes}
\label{sec:CGP}

Consider $x_n [k]$, a discrete time series on node $v_n$ in graph $G=(\V,\A)$, where $n$ indexes the nodes of the graph and $k$ indexes the time samples. Let $N$ be the total number of nodes and $K$ be the total number of time samples, and
\[
\x[k]=\begin{array}{cccc} \big( x_0[k] & x_1[k] &\ldots & x_{N-1}[k] \big)^T\end{array} \in \mathbb{C}^N
\]
represents the graph signal at time sample $k$.

We consider a Causal Graph Process (CGP) to be a discrete time series $\x[k]$ on a graph $G=(\V,\A)$ of the following form,
\begin{align}
\label{eq:CGPmodel}
\x[k] =& \w[k] + \sum\limits_{i=1}^{M}P_{i}(\A,\c)\x[k-i]& \nonumber\\
= &\w[k] + \sum\limits_{i=1}^{M}\left(\sum\limits_{j=0}^{i}c_{ij}\A^j\right)\x[k-i]&\\
= &\w[k] + (c_{10} \I+c_{11} \A)\x[k-1]& \nonumber\\
& + \left(c_{20} \I + c_{21} \A + c_{22} \A^2 \right) \x[k-2] +\ldots& \nonumber\\
& + \left(c_{M0} \I + \ldots + c_{MM}\A^M \right)\x[k-M], \nonumber
\end{align}
where $P_i(\A,\c)$ is a matrix polynomial in $\A$, $\w[k]$ is statistical noise, $c_{ij}$ are scalar polynomial coefficients, and
\[
\c=\begin{array}{cccccc}\big(c_{10} & c_{11} &\ldots & c_{ij} & \ldots & c_{MM}\big)^T\end{array}
\]
is a vector collecting all the $c_{ij}$'s.

Note that the CGP model does \emph{not} assume Markovianity in the nodes and edges of the graph adjacency matrix. Instead, the CGP is an autoregressive process (Markov) in the time series as in~\eqref{eq:CGPmodel} whose coefficients $P_i(\A,\c)$ are graph filters; thus, the CGP can incorporate the influence of many more (sometimes all)
other nodes in a single step.
Matrix polynomial $P_i(\A,\c)$ is at most of order $\min(i,N_\A)$, reflecting that $\x[k]$ cannot be influenced by more than $i$-th order network effects from $i$ time steps ago and in addition is limited (mathematically) by $N_\A$, the degree of the minimum polynomial of $\A$. Typically, we take the model order $M\ll N_{\A}$, and for the remainder of the paper assume this holds for sake of notational clarity.

This model captures the intuition that activity on the network travels at some fixed speed (one graph shift per sampling period), and thus the activity at the current time instant at a given network node cannot be affected by network effects of order higher than that speed allows.
In this way, the CGP model can be seen as generalizing the spatial dimension of the light cone~\cite{goerg_licors:_2012} to be on a discrete manifold rather than only on a lattice corresponding to uniformly sampled space.

The current parameterization of the CGP model in~\eqref{eq:CGPmodel} raises issues with identifiability. To address them, we assume that $P_1(\A,\c)\ne \alpha\I$ for $\alpha\in\Rbb$. Then, without loss of generality, we can let $c_{10}=0$ and $c_{11}=1$ so that $P_1(\A,\c)=\A$. To verify this, consider the full parameterization using $(\A',\c')$. We show that we can instead use the reduced parameterization $(\A,\c)$ with $P_1(\A,\c)=\A$ to represent the same process. First, we start with
\begin{equation}
\begin{aligned}
P_1(\A',\c')&=c_{10}'\I+c_{11}'\A'=\A=P_1(\A,\c)\\
\Rightarrow \A'&=(c_{11}')^{-1}(\A-c_{10}'\I).
\end{aligned}
\label{eq:CGP_param1}
\end{equation}
We can invert $c_{11}$, since by assumption $c_{11}'\ne 0$. Then consider the $i$-th polynomial,
\begin{align}
P_i(\A',\c')&=\sum\limits_{j=0}^{i}c_{ij}'(\A')^j=\sum\limits_{j=0}^{i}c_{ij}'(c_{11}')^{-j}(\A-c_{10}'\I)^j \nonumber\\
&=\sum\limits_{j=0}^{i}c_{ij}'(c_{11}')^{-j}\sum\limits_{k=0}^{j}\left(\!\begin{array}{c}\!j\! \\ \! k\!\end{array}\!\right)(-c_{10}')^{j-k}\A^k \nonumber\\
&=\sum\limits_{k=0}^{i}\sum\limits_{j=k}^{i}c_{ij}'(c_{11}')^{-j}\left(\!\begin{array}{c}\!j \! \\ \! k\!\end{array}\!\right)(-c_{10}')^{j-k}\A^k\\
&=\sum\limits_{k=0}^{i}c_{ik}\A^k =P_i(\A,\c), \nonumber
\label{eq:CGP_parami}
\end{align}
when we define
\[
c_{ik}=\sum\limits_{j=k}^{i}c_{ij}'(c_{11}')^{-j}\left(\!\begin{array}{c}\!j\! \\ \!k\!\end{array}\!\right)(-c_{10}')^{j-k}.
\]
In the remainder of this paper, we assume that $P_1(\A,\c)\ne \alpha\I$ and use the reduced parameterization with $c_{10}=0$ and $c_{11}=1$ so that $\c\in\Rbb^{n}$ where $n = (M-1)(M+4)/2$ to ensure that $\A$ and $\c$ are uniquely specified without ambiguity.

\section{Estimating Adjacency Matrices}
\label{sec:Estim}
Given a time series $\x(t)$ on graph $G=(\V,\A)$ with \emph{unknown} $\A$, we wish to estimate the adjacency matrix $\A$. We assume the data follows the CGP model~\eqref{eq:CGPmodel}. A first approach to its estimation can be formulated as the following optimization problem,
\begin{equation}
\label{eq:OptNonconvex2}
\begin{aligned}
(\A,\c)=\underset{\A,\c}{\argmin} &\; \frac{1}{2} \sum\limits_{k=M}^{K-1}\left\| \x[k] - \sum\limits_{i=1}^{M}P_{i}(\A,\c)\x[k-i] \right\|_2^2 \\
&+\lambda_1 \|\vc(\A)\|_1 +\lambda_2 \|\c\|_1,
\end{aligned}
\end{equation}
where $\vc(\A)$ stacks the columns of the matrix $\A$.

In equation~(\ref{eq:OptNonconvex2}), the first term in the right hand side models $\x[k]$ by the CGP model~\eqref{eq:CGPmodel} in section~\ref{sec:CGP}, the regularizing term $\lambda_1 \|\vc(\A)\|_1$ promotes sparsity of the estimated adjacency matrix, and the term $\lambda_2 \|\c\|_1$ promotes sparsity in the matrix polynomial coefficients.
Regularizing $\c$ corresponds to performing autoregressive model order selection.
If the true model has $P_i(\A,\c)\!=\!\0\;\forall j\le i\le M$ for some $0<j<M$, then regularization of $\c$ encourages the corresponding values to be $\0$.
The matrix polynomial in the first term makes this problem nonconvex.
That is, using a convex optimization based approach to solve~(\ref{eq:OptNonconvex2}) directly may result in finding a solution $(\widehat{\A},\widehat{\c})$, minimizing the objective function locally, that is not near the true globally minimizing~$(\A,\c)$.

Instead, we break this estimation down into three separate, more tractable steps:

\begin{cenum}
\item Solve for $\R_i=P_i (\A,\c)$
\item Recover the structure of $\A$
\item Estimate $c_{ij}$
\end{cenum}

\subsection{Solving for $P_i(\A,\c)$}
\label{subsec:SolvePi}
As previously stated, the graph filters $P_i(\A,\c)$ are polynomials of A and are thus shift-invariant and must mutually commute. Then their commutator
\begin{align*}
[P_i (\A,\c),\;&P_j (\A,\c)]=\\
&P_i (\A,\c) P_j (\A,\c) - P_j (\A,\c) P_i (\A,\c) = 0 \; \; \forall i,j.
\end{align*}
Let $\R_i=P_i(\A,\c)$, $\R=(\R_1,\ldots,\R_M)$, and $\widehat{\R}_i$ be the estimate of $P_i(\A,\c)$. This leads to the optimization problem,

\begin{equation}
\label{eq:OptP}
\begin{aligned}
\widehat{\R}=&\underset{\R}{\argmin}\;  \frac{1}{2} \sum\limits_{k=M}^{K-1} \left\| \x[k] - \sum\limits_{i=1}^{M}\R_i \x[k-i] \right\|_2^2 \\
+&\lambda_1 \left\|\vc(\R_1 )\right\|_1 +\lambda_3 \sum\limits_{i\ne j}\left\|[\R_i,\R_j]\right\|_F^2.
\end{aligned}
\end{equation}
While this is still a non-convex problem, it is multi-convex. That is, when $\R_{-i}=\{ \R_j\, : \, j\ne i \}$ (all $\R_j$ except for $\R_i$) are held constant, the optimization is convex in $\R_i$. This naturally leads to block coordinate descent as a solution,

\begin{equation}
\label{eq:OptRi}
\begin{aligned}
\widehat{\R}_i=\underset{\R_i}{\argmin} &\; \frac{1}{2} \sum\limits_{k=M}^{K-1} \left\| \x[k] - \sum\limits_{j=1}^{M}\R_j \x[k-j] \right\|_2^2 \\
+&\lambda_1 \|\vc(\R_1 )\|_1 +\lambda_3\sum\limits_{j\ne i}\left\|[\R_i,\R_j]\right\|_F^2.
\end{aligned}
\end{equation}
Each of these sub-problems for estimating $\R_i$ in a single sweep of estimating $\R$ is formulated as an $\ell_1$-regularized least-squares problem that can be solved using standard gradient-based methods~\cite{figueiredo_gradient_2007}. We compute a rough estimate of the computational cost for solving~\eqref{eq:OptRi}. Assume that finding the gradient is the most costly step in the optimization, and naively estimate matrix-matrix products to take $O(N^3)$ operations; then the optimization has worst-case complexity $O(M^2 N^3 + KMN^2 )$ incurred from minimizing over $M$ separate blocks. The cost of the $i$-th problem is dominated in each iteration by $K-M$ matrix-vector products $\R_i \x[k-i]$ with worst case total complexity $O((K-M) N^2 )$ and by matrix-matrix products $\R_i\R_j$ for $j\ne i$ with worst case complexity $O((M-1)N^3)$. We now improve these cost estimates by noting that, due to the sparsity in the $\R_i$ matrices, the factor of $O(MN^3)$ resulting from matrix multiplications can be reduced to $O(MS^2_{N})$ when implemented using sparse matrix multiplications, where $S_{N}$ is the sparsity of the $\R_i$ (i.e., $\max_{i}\|\R_i\|_0 = S_{N}$). In addition, the sparse matrix-vector products can also be reduced to $O(MS_{N})$. Then, the total complexity is better estimated to be $O(MS^2_{N})$, which scales more amenably for large data applications.
\subsection{Recovering $\A$}
\label{subsec:RecoverA}

After obtaining estimates $\widehat{\R}_i$, we find an estimate for $\A$. One approach is to take $\widehat{\A} = \widehat{\R}_1$. This appears to ignore the information from the remaining $\widehat{\R}_i$. However, this information is taken into account during the iterations when solving for $\widehat{\R}$, especially if we begin one new sweep to estimate $\R_1$ using~(\ref{eq:OptRi}) with $i=1$. A second approach is also possible, explicitly using all the $\widehat{\R}_i$ together to find $\A$,
\begin{equation}
\label{eq:findA2}
\begin{aligned}
\widehat{\A}=\underset{\A}{\argmin}&\;\left\|\widehat{\R}_1-\A\right\|_2^2+\lambda_1\|\vc(\A)\|_1\\
&+\lambda_3\sum\limits_{i=2}^{M}\left\|\left[\A,\widehat{\R}_i \right]\right\|_F^2.
\end{aligned}
\end{equation}
This can be seen as similar to running one additional step further in the block coordinate descent to find $\widehat{\R}_1$, except that this approach does not explicitly use the data. This has worst-case complexity $O(MN^3)$ dominated by matrix-matrix products $\A\R_i$ for $i\ne 1$. However, when matrices $\A$ and $\R_i$ are sparse, we again have reduced complexity of $O(MS^2_{N})$.

\subsection{Estimating $\c$}
\label{subsec:Estimcij}
We can estimate $c_{ij}$ in one of two ways: we can estimate $\c$ either from $\widehat{\A}$ and $\widehat{\R}_i$ or from $\widehat{\A}$ and the data $\X$.

To estimate $c_{ij}$ from $\widehat{\A}$ and $\widehat{\R}$, we set up the optimization,
\begin{equation}
\label{eq:FindCSpec}
\widehat{\c}_i = \underset{\c_i}{\argmin}\frac{1}{2}\left\|\vc\left(\widehat{\R}_i\right) - \Q_i\c_i \right\|_2^2+\lambda_2 \|\c_i \|_1,
\end{equation}
where\par
{\small \qquad $\Q_i = \begin{array}{cccc}\Big(\vc(\I) & \vc(\widehat{\A}) & \ldots &\vc(\widehat{\A}^i) \Big)\end{array}$,\par
\qquad $\c_i =\begin{array}{cccc} \big(c_{i0} & c_{i1} & \ldots & c_{ii} \big)^T\end{array}.
$\par
}Alternatively, to estimate $c_{ij}$ from $\widehat{\A}$ and the data $\X$, we can use the optimization,
\begin{equation}
\label{eq:FindC2Spec}
\widehat{\c} = \underset{\c}{\argmin}\;\frac{1}{2}\left\|\Y(\what\A) - \B(\what\A)\c \right\|_F^2+\lambda_2 \|\c \|_1
\end{equation}
where $\Y(\what\A)=\vc\left(\X_M-\what\A\X_{M-1}\right)$,\\
$\begin{aligned}&\B(\what\A)\!=\!\!\!\begin{array}{ccccc}\Big(\!
\vc\left(\X_{M-2}\right) \!\! & \!\!...\!\! & \!\! \vc\left(\widehat{\A}^j \X_{M-i}\right) \!\! & \!\! ... \!\! & \!\! \vc\left(\widehat{\A}^M \X_{0}\right)\!
\Big),\end{array}\\
&\textrm{and }\X_m =\left(\!\begin{array}{cccc}\x[m] & \x[m+1] & ... & \x[m+K-M-1]\end{array}\!\right),\end{aligned}$ where in $\B(\what\A)$ the $i$ and $j$ indices increment in correspondence to the indexing of $c_{ij}$ in $\c$. That is, first set $i=2$; then $j$ is incremented from $0,\ldots,i$, and $i$ is incremented, and this repeats until $(i,j)=(M,M)$.
Then~\eqref{eq:FindC2Spec} can also be solved using standard $\ell_1$-regularized least squares methods with worst-case complexity $O(M^4)$ per iteration, dominated by computing (dense) matrix-vector products with $\B^\top\B\in\Rbb^{n \times n}$ and $\B^\top\Y\in \Rbb^{n}$, with $n=(M-1)(M+4)/2=O(M^2)$. While this may seem daunting, we note that $M$, the lag order of the autoregressive process, is usually much lower than the total number of time samples $K$, so that the optimization to find $\c$ is unlikely to be the bottleneck in the overall model estimation.

\subsection{Base Estimation Algorithm}
\label{subsec:GenEst}
The methods discussed so far can be interpreted as assuming that 1) the process is a linear autoregressive process driven by white Gaussian noise and 2) the elements in parameters $\A$ and $\c$ a priori follow zero-mean Laplace distributions. Under these assumptions, the objective function in~(\ref{eq:OptNonconvex2}) approximately corresponds to the log posterior density and its optimization to an approximate maximum a posteriori (MAP) estimate.

This framework can be extended to estimate more general autoregressive processes, such as those with a non-Gaussian noise model and certain forms of nonlinear dependence of the current state on past values of the state. In this case, we can formulate the general optimization as\par
{\vspace{-1em}\small\begin{equation}
\label{eq:OptNonconvex}
\begin{aligned}
(\what{\A},\what{\c})\!=\!\underset{\A,\c}{\argmin} &f_1\Big(\!\vc(\X_M), \vc\Big(\sum\limits_{i \!=\! 1}^M \! P_i(\A,\c)\X_{M \!-\! i} \Big) \!\Big) \\
&+g_1 \big(\A\big) +g_2 (\c),
\end{aligned}
\end{equation}
}where matrices $\X_{m}$ are defined under~\eqref{eq:FindC2Spec},  $f_1(\cdot,\cdot)$ is a loss function that corresponds to a log-likelihood function dictated by the noise model, and $g_1(\cdot)$ and $g_2(\cdot)$ are regularization functions (usually convex norms) that correspond to log-prior distributions imposed on the parameters and are dictated by modeling assumptions. Again, the matrix polynomials $P_i(\A,\c)$ introduce nonconvexity, so similarly as before, we can separate the estimation into three steps to reduce complexity. This leads to analogous formulations for the optimization problems~\eqref{eq:OptNonconvex2}-\eqref{eq:FindC2Spec} that we omit for brevity and clarity. In the remainder of this paper, we refer to the specific formulations given in~\eqref{eq:OptNonconvex2}-\eqref{eq:FindC2Spec}.

Algorithm~\ref{alg:AlgBasic} summarizes the 3-step algorithm outlined in this section for obtaining estimates $\widehat{\A}$ and $\widehat{\c}$ for the adjacency matrix and filter coefficients; it is a more efficient and well-behaved alternative to directly using~(\ref{eq:OptNonconvex}).

\begin{algorithm}
\caption{Base estimation algorithm}
\label{alg:AlgBasic}
\begin{algorithmic}[1]
\State Initialize, $t=0$, $\widehat{\R}^{(t)}=\0$
\While{$\widehat{\R}^{(t)}$ not converged}
	\For{$i=1:M$}
		\State Find $\widehat{\R}_i^{(t)}$ with fixed $\widehat{\R}_{<i}^{(t)}$, $\widehat{\R}_{>i}^{(t-1)}$ using~(\ref{eq:OptRi}).
	\EndFor
	\State $t\gets t+1$
\EndWhile
\State Set $\widehat{\A}=\widehat{\R}_1^{(t)}$ or estimate $\widehat{\A}$ from $\widehat{\R}^{(t)}$ using~(\ref{eq:findA2}).
\State Solve for $\widehat{\c}$ from $\X$, $\widehat{\A}$ using~(\ref{eq:FindCSpec}) or from $\X$, $\widehat{\R}$ using~(\ref{eq:FindC2Spec}).
\end{algorithmic}
\end{algorithm}

We call this 3-step procedure the base algorithm. In Algorithm~\ref{alg:AlgBasic}, superscripts denote the iteration number, $\widehat{\R}_{<i}^{(t)}$ denotes $\{\widehat{\R}_j^{(t)} : j<i\}$ and likewise $\widehat{\R}_{>i}^{(t)}$ denotes $\{\widehat{\R}_j^{(t)} : j>i\}$.

\subsection{Simplified Estimation Algorithm}
\label{subsec:SimpEst}
The base algorithm can still be moderately expensive to evaluate computationally when scaling to larger problems and is difficult to analyze theoretically, mainly due to the nonconvexity of the commutativity-enforcing term.
For further ease of computation and analysis, we consider a simplified version of the base algorithm in which the commutativity term of the optimization problem~\eqref{eq:OptRi} is removed,
\begin{equation}
\begin{aligned}
\widehat{\R}=\underset{\R}{\argmin} \; &f_{1}\Big(\vc(\X_M), \vc\Big(\sum\limits_{i=1}^{M} \R_i\X_{M-i}\Big) \Big)\\ &+\sum\limits_{i=1}^M g_{1i}\big(\R\big).
\end{aligned}
\label{eq:OptRSimp}
\end{equation}
Note that we have recombined the optimization to be joint over all the $\R_i$, since this is convex without the commutativity term. This can be followed by the same steps to find $\what{\A}$ and $\what{\c}$ as in the base algorithm, taking $\what{\A}=\what{\R}_1$ or as the solution to~\eqref{eq:findA2}, and then solving~\eqref{eq:FindCSpec} or~\eqref{eq:FindC2Spec} for $\what{\c}$. We call this the Simplified Algorithm, described in Algorithm~\ref{alg:AlgSimp}. Using standard solvers for the estimation in~\eqref{eq:OptRSimp} with $\ell_2$ and $\ell_1$ norms as seen before~\cite{figueiredo_gradient_2007}, the worst-case cost is $O(M(K-M)N^2 )$, dominated by computing the matrix-vector product $\R_i \x[k-i]$ for $i=1,\ldots, M$ and for $k=M,\ldots,K$. Again, by implementing sparse matrix-vector products, we can reduce the complexity to $O(M(K-M)S_{MN})$ in the best case.

\begin{algorithm}
\caption{Simplified estimation algorithm}
\label{alg:AlgSimp}
\begin{algorithmic}[1]
\State Initialize $\widehat{\R}=\0$
\State Estimate $\what{\R}$ using~\eqref{eq:OptRSimp}
\State Set $\widehat{\A}=\widehat{\R}_1$ or estimate $\widehat{\A}$ from $\widehat{\R}$ using~(\ref{eq:findA2}).
\State Solve for $\widehat{\c}$ from $\X$, $\widehat{\A}$ using~(\ref{eq:FindCSpec}) or from $\X$, $\widehat{\R}$ using~(\ref{eq:FindC2Spec}).
\end{algorithmic}
\end{algorithm}
\subsection{Extended Estimation Algorithm}
\label{subsec:ExtEst}

As an extension of the base algorithm, we can also choose the estimated matrix $\widehat{\A}$ and filter coefficients $\widehat{\c}$ to initialize the direct approach of using~(\ref{eq:OptNonconvex}).
Starting from these initial estimates, we may find better local minima than with initializations at $\A=\0$ and $\c=\0$ or at random estimates.
We call this procedure the extended algorithm, summarized in Algorithm~\ref{alg:AlgExt}. 

\begin{algorithm}
\caption{Extended estimation algorithm}
\label{alg:AlgExt}
\begin{algorithmic}[1]
\State Initialize $\A^{(0)}=\0$ and $\c^{(0)}=\0$.
\State Estimate $\A^{(0)}$, $\c^{(0)}$ using basic algorithm.
\State Find local minimum $\widehat{\A}$, $\widehat{\c}$ using initialization $\A^{(0)}$, $\c^{(0)}$ from nonconvex problem~(\ref{eq:OptNonconvex}) using convex methods to find a local optimum.
\end{algorithmic}
\end{algorithm}
\section{Convergence of Estimation}
\label{sec:Conv}

In this section, we discuss the convergence of the basic, extended, and simplified algorithms described above. Convergence of the Base Algorithm, Algorithm~\ref{alg:AlgBasic}, and of the Extended Algorithm, Algorithm~\ref{alg:AlgExt} follow by direct application of the results available in literature, as discussed in Sections~\ref{subsec:ConvB} and~\ref{subsec:ConvExt}. Performance of the Simplified Algorithm, Algorithm~\ref{alg:AlgSimp}, is proven in Section~\ref{subsec:ConvSimp}.

\subsection{Base Estimation Algorithm}
\label{subsec:ConvB}

In estimating $\A$ and $\c$, as illustrated in equations~(\ref{eq:findA2}) and~(\ref{eq:FindC2Spec}), the forms of the optimization problems are well studied when choosing $\ell_2$ and $\ell_1$ norms as loss and regularization functions~\cite{tibshirani_regression_1996,basu_regularized_2015}. However, using these same norms, step~2 (the while loop) of the Base Algorithm is a nonconvex optimization. Hence, we would like to ensure that step 2 converges.

When using block coordinate descent for general functions (i.e., repeatedly choosing one block of coordinate directions in which to optimize while holding all other blocks constant), neither the solution nor the objective function values are guaranteed to converge to a global or even a local minimum. However, borrowing results from~\cite{tseng_convergence_2001}, under some mild assumptions, using block coordinate descent to estimate $\R_i$ will converge. In fact, in equation~(\ref{eq:OptNonconvex}), if we assume the objective function to be continuous and to have convex, compact sublevel sets in each coordinate block $\R_i$ (for example, if the functions for $f_1$, $g_{1}$, and $g_2$ are the $\ell_2$, $\ell_1$, and $\ell_1$ norms respectively as in equation~(\ref{eq:OptNonconvex2})), then the block coordinate descent will converge.

\subsection{Extended Estimation Algorithm}
\label{subsec:ConvExt}

Now we discuss the convergence of the extended estimation algorithm described in Algorithm~\ref{alg:AlgExt} of section~\ref{subsec:ExtEst} assuming that in step 1 of Algorithm 3, the Base Algorithm has converged to an initial point $(\A^{(0)},\c^{(0)})$.
We assume that the objective function $F=f_1+g_1+g_2$ in equation~(\ref{eq:OptNonconvex}) has compact sublevel sets and is bounded below. Then an iterative (sub-)gradient method with appropriately chosen step sizes that produces updates of $(\A^{(t+1)},\c^{(t+1)})$ such that $F(\A^{(t+1)},\c^{(t+1)})\le F(\A^{(t)},\c^{(t)})$, may converge to a local optimum (which is not necessarily the global optimum). If the functions $f_1$, $g_1$, and $g_2$ are the $\ell_2$, $\ell_1$, and $\ell_1$ norms as in equation~(\ref{eq:OptNonconvex2}), these conditions are satisfied and the algorithm converges to a local optimum.
\subsection{Simplified Estimation Algorithm}
\label{subsec:ConvSimp}

Here, we consider the convergence of the Simplified Algorithm, Algorithm~\ref{alg:AlgSimp}. We state in this section formally the underlying assumptions and the results; we also outline the main arguments underlying the proofs. The proofs themselves are detailed in Appendices~\ref{app:pf_CGP_estim1}-\ref{app:CGP1}. We consider the theoretical performance guarantees provided by Algorithm~\ref{alg:AlgSimp} of the simplified estimate $(\what{\A},\what{\c})$ when using $\ell_1$ regularized least squares to estimate $\what{\R}$,
\begin{equation}
\begin{aligned}
\widehat{\R}=\underset{\R}{\argmin} \; &\frac{1}{\sigma_u}\sum\limits_{k=M}^{K-1}\left\|\x[k]-\sum\limits_{i=1}^{M}\R_i\x[k-i]\right\|_2^2 \\
&+\lambda_1 \|\vc(\R)\|_1,
\end{aligned}
\label{eq:OptRSimpSpecific}
\end{equation}
which is a special case of~\eqref{eq:OptRSimp}, where $\sigma_u=\|\Ebb[\w_k\w_k^\top]\|$, and then using $\what{\A}=\what{\R}_1$ and the optimization~\eqref{eq:FindC2Spec} to find $\what{\c}$. Dividing by $\sigma_u$ makes the estimation unitless, although in practice we might incorporate its value into the $\lambda_1$ parameter.

Our error metric of interest will be:
\begin{equation}
\label{eq:CGP_error_metric}
\begin{aligned}
	\epsilon
	=&\Ebb\left[ \frac{1}{N}\left\|\x[k]-f(\widehat{\A},\widehat{\c},\X_{k-1}') \right\|_2^2 \right]\\
	&-\Ebb\left[ \frac{1}{N}\Big\|\x[k]-f(\A,\c,\X_{k-1}') \Big\|_2^2 \right],
\end{aligned}
\end{equation}
where\par
{\centering
$\X_{k\!-\!1}' \!=\! \left(\!\begin{array}{cccc}\x[k \!-\!1]^\top \!&\! \x[k \!-\! 2]^\top \!&\! \ldots \!&\! \x[k \!-\! M]^\top \end{array}\!\right)^\top \!\in\! \Rbb^{MN}$
\par}
\noindent and $f(\A,\c,\X_{k-1}')=\sum\limits_{i=1}^{M}P_i(\A,\c)\x[k-i]$. This error $\epsilon$ is the average excess prediction risk, the difference between the error of estimating $x[k]$ by the estimated CGP (first term in~\eqref{eq:CGP_error_metric}) and the variance of the noise $\w[k]$ in the CGP given in~\eqref{eq:CGPmodel} (second term in~\eqref{eq:CGP_error_metric}).
The expectations in~\eqref{eq:CGP_error_metric} are taken over a new sample
\begin{align}
\z_k=\Big(\x[k]^\top \; \X_{k-1}'^\top \Big)^\top \in \Rbb^{(M+1)N}
\label{eq:zk}
\end{align}
drawn independently of the samples used to estimate $\left(\widehat{\A},\widehat{\c}\right)$.
We now state the assumptions underlying the main results.
\subsubsection{Assumptions}
\label{subsubsec:assump}

First, we list the assumptions we make about the true process in order to derive our performance guarantees:
\begin{enumerate}[label=\textbf{(A\arabic*)}]
	\item \label{asu:CGP_model} The CGP model class~\eqref{eq:CGPmodel} is accurate:\\
	$\Ebb[\x[k]\,|\,\X_{k-1}']=f(\A,\c,\X_{k-1}')$.
	\item \label{asu:CGP_noise} The noise process is uncorrelated with the CGP and its own past values:\\
	$\Ebb\left[\x[j]\w[k]^\top \right]\!=\!\0$ and $\Ebb\left[\w[j]\w[k]^\top \right]\!=\!\0$ for $j < k$. Further, the noise sequence $\{\w[k]\}$ is i.i.d. multivariate Gaussian with distribution $\w[k]\sim\Ncal(\0,\boldsymbol{\Sigma}_\w)$, and there exists $\sigma_\ell$ such that $0<\sigma_\ell \le \left\|\boldsymbol{\Sigma}_\w^{-1}\right\|_2^{-1}$, the singular values of $\boldsymbol{\Sigma}_\w$ are strictly bounded away from $0$. Then we can represent the condition number of $\boldsymbol{\Sigma}_\w$ as $\sigma_u/\sigma_\ell$, where $\sigma_u$ is given in~\eqref{eq:OptRSimpSpecific}.
	\item \label{asu:CGP_steady} The CGP is stationary and is already in steady state when we begin sampling. Under this assumption, the marginal distributions and expectations are $\Ebb[\x[k]]=\0$, $\boldsymbol{\Sigma}_0\overset{\Delta}{=}\Ebb\left[\x[k]\;\x[k]^\top \right]$, and $\boldsymbol{\Sigma}\overset{\Delta}{=}\Ebb\left[\z_k \z_k^\top \right]$ where $\z_k$ is defined as in~\eqref{eq:zk}.
	\item \label{asu:CGP_sigma_abs_sum} The stationary correlation matrices of the process $\x[k]$ are absolutely summable in induced norm:\par	
	{\centering $\displaystyle \sum\limits_{i=-\infty}^{\infty}\left\|\Ebb\left[\x[k]\x[k-i]^\top \right]\right\|_2=G<\infty.$ \par} This is a slightly stronger condition than stability (see proof of Lemma~\ref{lem:CGP_estim1} in appendix~\ref{app:pf_CGP_estim1}). 
	\item \label{asu:CGP_sparse} The true adjacency matrix and filter coefficients are sparse and bounded:\par
	{\small\centering
	$\left\|\left(\!\begin{array}{cccc}\A&P_2(\A,\c)&\ldots&P_M(\A,\c)\end{array}\!\right)\right\|_0\le S_{MN} \ll MN^2$
	\par}and $\max_{1\le i\le M}(\|\A^i\|_2) \le L$, $\max_{1\le i\le M}(\|\A^i\|_0) \le t_N \ll N^2$, and $\|c\|_0\le s_M$ and $\|\c\|_1 \le \rho$, where the matrix $\ell_0-$``norm'' $\|\A\|_0$ is the total number of nonzero entries in matrix $\A$. The quantities $S_{MN}$ and $s_{M}$ may grow with $M$ and $N$, and $t_N$ may grow with $N$.
	\item \label{asu:CGP_L_rho} The following holds for the bounds $L$ and $\rho$ in~\ref{asu:CGP_sparse}: $(1+L)(1+\rho)=Q\le 2$. This is also a slightly stronger condition than stability (see proof of Lemma~\ref{lem:CGP_estim1} in Appendix~\ref{app:pf_CGP_estim1}).
	\item \label{asu:CGP_samples} The sample size $K$ is large enough relative to the ``stability'' of the process, the log of the network size, and the sparsity. Also the autoregressive model order $M$ is low relative to the length of the sample size $K$:\par
	{\centering $\begin{aligned}T&=K-M\ge C \omega^2\,S_{MN}\,(\log M+\log N) \\ M &\lesssim o(\log K)\end{aligned}$ \par} for some universal constant $C>0$ and $\omega=\frac{\sigma_u}{\sigma_\ell }\frac{\ Q^2}{\mu_{\min}(\wtil{\Acal})}$, where $\omega$ and $\mu_{\min}(\wtil{\Acal})$ are related to measures of ``stability'' of the process~\cite{basu_regularized_2015}, and their explicit forms are given in appendix~\ref{app:pf_CGP_estim1} in equations~\eqref{eq:Atilde} and~\eqref{eq:omega}, and $Q$ is given in~\ref{asu:CGP_L_rho}.
	\item \label{asu:CGP_BA_min} The matrix $\Ebb[\B(\A)^\top\B(\A)]$ is invertible, or alternatively, its minimum singular value is strictly positive: $\left\|\left(\Ebb[\B(\A)^{\top}\B(\A)]\right)^{-1}\right\|_2^{-1}\ge \kappa_\B NT >0$.
\end{enumerate}
We point out that assumptions~\ref{asu:CGP_model}-\ref{asu:CGP_steady},~\ref{asu:CGP_samples}, and~\ref{asu:CGP_BA_min} correspond to fairly standard assumptions in stationary time series analysis and studying performance of parameter estimation. \ref{asu:CGP_samples} assumes enough samples to do meaningful estimation, which is standard in high-dimensional estimation. In this assumption, the minimum number of samples is linked to the properties of the process. Assumption~\ref{asu:CGP_BA_min} makes sure that $\c$ is identifiable when given the true value of $\A$.

As noted~\ref{asu:CGP_sigma_abs_sum} and~\ref{asu:CGP_L_rho} are slightly stronger than ``stability.'' This is because these assumptions are not necessarily implied by stationarity. We note that~\ref{asu:CGP_sparse} is perhaps the most restrictive assumption, as it imposes explicit sparsity conditions on the polynomials $P_i(\A,\c)$ and vector $\c$. In network science terms, this assumption roughly corresponds to graphs with relatively longer node-to-node paths, so that higher order powers of $\A$ are not all dense. However, this assumption has the same flavor as the explicit sparsity assumptions made in other sparse estimation work. It would be interesting to relax this assumption to recent notions of approximate sparsity rather than exact sparsity, and that could be the direction of future work.

\subsubsection{Theoretical Performance}
\label{subsubsec:guarantee}

Here, we present the main guarantee and a brief sketch for its proof, providing several lemmas of intermediate results used in the main result. The full proof is presented in Appendices~\ref{app:pf_CGP_estim1}-\ref{app:CGP1}.

\begin{thm}[Main Result]
	\label{thm:main}
	For any $0<\beta<\nu<1/2$, and some universal constant $d_1$, assumptions~\ref{asu:CGP_model}-\ref{asu:CGP_BA_min} are sufficient for the error $\epsilon$ in~\eqref{eq:CGP_error_metric} to satisfy
	{\small
		\begin{equation}
		\label{eq:CGP_theorem}
		\begin{aligned}
		\epsilon \le \left(\delta_\A \left(1+ (\rho+\delta_\c) \what L_M(\delta_\A) \right) + (1+L) \delta_\c\right)^2\tr(\boldsymbol{\Sigma}_0)/N
		\end{aligned}
		\end{equation}
	}with probability at least $1-\varepsilon_\A-\varepsilon_\c$, where
	\[\what L_M(\delta)=\max_{1\le i \le M}\; \frac{(L+\delta)^i-L^i}{\delta}\]
	\begin{equation}
	\begin{aligned}
	\varepsilon_\A &\sim d_1 \exp\{-O(K)\}\\
	\varepsilon_\c &\sim 2\exp\{ -O(K^{1-2\nu})\}\\
	\delta_\A &=O\left(\sqrt{\log N/K}\right)\\
	\delta_\c &=O\left(\sqrt{\log N/K^{2(\nu-\beta)}}\right).
	\end{aligned}
	\end{equation}
\end{thm}
The exact expressions for $\epsilon_\A$, $\epsilon_\c$, $\delta_\A$, and $\delta_\c$ can be found in the appendices.
This theorem states that, as the number of nodes $N$ and the number of time observations $K$ grow, with high probability, the average excess prediction risk of the simplified estimate is not too large. The dependence of $\delta$ on $K$ and $L$ are through $T=K-M$ and $Q$, respectively, where $L$ is defined in~\ref{asu:CGP_sparse}, and $Q$ is defined in~\ref{asu:CGP_L_rho}. The AR order $M$ and network size $N$ may also grow, as long as they grow slowly enough with respect to $K$. Note that, for large $K$, we have $\delta \ll L$, and the factor $\what{L}_M(\delta)=O(ML^{M-1})$. The full proof of Theorem~\ref{thm:main} is in Appendix~\ref{app:CGP1}, but we provide a brief overview here.

Before we outline the proof of the theorem, we first present two intermediate results.
These two results use sparse vector autoregression estimation results to show that $\|\what\A-\A\|_2$ and $\|\what\c-\c\|_1$ are small with high probability. Then, with small errors in estimating $\A$ and $\c$, we can demonstrate that the error $\epsilon$ is also small.

\begin{lem}
	\label{lem:CGP_estim1}
	Assume~\ref{asu:CGP_model} that $\x[k]$ is generated according to the CGP model with $\A$ satisfying \ref{asu:CGP_sparse}.
	Suppose~\ref{asu:CGP_samples} that the sample size $K$ is large enough. Let $d_i \!>\! 0$ with $i \!=\! 1,...,3$ be universal constants, and let {\small $g(Q)=d_3 \left( 1+\frac{1+Q^2}{(2-Q)^2}\right)$} and $\ell_{MNK}=\sqrt{(\log M+\log N)/K}$. With $\varepsilon_\A=d_1 \exp\{-d_2 K /\omega^2\}$, $\lambda_1 \!=\! 4 g(Q) \ell_{MNK}$, and $\what\R=(\what\R_1,\ldots,\what\R_M)$ the solution to~\eqref{eq:OptRSimpSpecific}, with probability at least $1-\varepsilon_\A$, $\what\A=\what\R_1$ satisfies the inequalities\par
	{\centering \small $\begin{aligned}
		\|\what{\A}-\A\|_1 &\le 256 S_{MN} \ell_{MNK} Q^2 g(Q)  \sigma_u/\sigma_\ell \\
		\|\what{\A}-\A\|_2 \le \|\what{\A}-\A\|_F &\le \delta_{\A}\overset{\Delta}{=} 64 \sqrt{S_{MN}} \ell_{MNK} Q^2 g(Q) \sigma_u/\sigma_\ell\end{aligned}$\par
	}
	\noindent
	where $\omega=\frac{\sigma_u}{\sigma_\ell }\frac{\ Q^2}{\mu_{\min}(\wtil{\Acal})}$ as defined in~\ref{asu:CGP_samples}.		
\end{lem}
This lemma states that, with the appropriate choice of $\lambda_1$, the assumptions~\ref{asu:CGP_model}-\ref{asu:CGP_BA_min} are sufficient to allow good estimation of $\A$ with high probability. That is, for each increase of sample size $K$, we choose the optimal value of the regularization parameter $\lambda_1$, and this choice yields consistency with high probability.

\begin{lem}
	\label{lem:CGP_c_body}
	Suppose that assumptions~\ref{asu:CGP_model}-\ref{asu:CGP_BA_min} hold. Then for any $0<\beta<\nu <1/2$, and sparsity $s_M\le d_4 K/\log n $ where $d_4>0$ is a universal constant, and $\lambda_2 \ge q_1= \Theta(N\sqrt{\log N}K^{1-\beta})$, the solution to~\eqref{eq:FindC2Spec} satisfies
	\[
	P\left( \|\c-\what\c\|_1 \ge \delta_\c\right)\le \varepsilon_\c
	\]
	where
	\begin{align*}
	\delta_\c &=O\left(\sqrt{\log N/K^{2(\nu-\beta)}}\right)\\
	\varepsilon_\c &\sim 2\exp\{-O(K^{1-2\nu})\}.
	\end{align*}
\end{lem}
In a similar vein as Lemma~\ref{lem:CGP_estim1}, this lemma states that, for appropriate choice of $\lambda_2$, the assumptions~\ref{asu:CGP_model}-\ref{asu:CGP_BA_min} are sufficient to yield a good estimate of $\c$ with high probability.

\begin{proof}[Proof Overview for Theorem~\ref{thm:main}]
	Lemma~\ref{lem:CGP_estim1} shows that we can achieve good performance in estimating $\A$ with high probability. With considerable effort and several additional insights, we show that good performance in estimating $\A$, translates into good estimation of $\c$ with high probability in Lemma~\ref{lem:CGP_c_body}.
	
	Then, small errors $\|\what\A-\A\|_2 \le\delta_\A$ and $\|\what\c-\c\|_1\le\delta_\c$ naturally lead to small prediction errors $\epsilon$. Concretely, we proceed with some algebra, showing that
	\begin{align*}
	\epsilon&=\frac{1}{N}\Ebb\left[ \left\|f\left(\what{\A},\what\c,\X_{k-1}'\right)-f\left(\A,\c,\X_{k-1}'\right) \right\|_2^2 \right]\\
	&\le \frac{1}{N}\Ebb \begin{aligned}[t]&\left[ \left(\|f(\what\A,\what\c,\X_{k-1}') \!-\! f(\A,\what\c,\X_{k-1}') \|_2 \right.\right. \\
	&\quad +\! \left.\left.\vphantom{\what\A} \|f(\A,\what\c,\X_{k-1}')\!-\!f(\A,\c,\X_{k-1}')\|_2\right)^2\right].\end{aligned}
	\end{align*}
	Given small estimation errors, we can bound these two norms separately. The first can be bounded using both Lemmas~\ref{lem:CGP_estim1} and~\ref{lem:CGP_c_body}, and the second can be bounded with Lemma~\ref{lem:CGP_c_body}. Applying the union bound, we arrive at our result. Again, the full detailed proof is in the appendices~\ref{app:pf_CGP_estim1}-\ref{app:CGP1}.
\end{proof}


\section{Experiments}
\label{sec:Exp}
We test our algorithms on two types of datasets, a real temperature sensor network time series (with $N=150$ and $K=365$) and a synthetically generated time series (with varying $N$ and $K$). With the temperature dataset, we compare the performance of the different algorithms (1, 2, and 3) for estimating the CGP model~\eqref{eq:CGPmodel} against that of the sparse vector autoregresssive Markov random field model~\eqref{eq:SVAR} labeled as MRF, where $f_1(\x,\y)=\frac{1}{2}\|\x-\y\|_2^2$ and $g_1(\x)=\lambda_1\|\x\|_1$; with the synthetic dataset, we compare the estimates of the model parameters to the ground truth graph used to generate the data for Algorithm 2, and we evaluate the performance of the prediction by averaging through Monte Carlo simulations to empirically test Theorem~\ref{thm:main}.

To solve the regularized least squares iterations for estimating the CGP matrices, we used a fast implementation of proximal quasi-Newton optimization for $\ell_1$ regularized problems~\cite{schmidt_fast_2007}. To estimate the MRF matrices, we used an accelerated proximal gradient descent algorithm~\cite{odonoghue_adaptive_2013} to estimate the SVAR matrix coefficients from equation~\eqref{eq:SVAR} since the code used in~\cite{bolstad_causal_2011} is not tested for large graphs.

\subsection{Temperature Data}
\label{subsec:Temperature}

The temperature dataset is a collection of daily average temperature measurements taken over $365$ days at $150$ cities around the continental United States~\cite{_national_2011}.
The time series $\x_i=\left(x_i[0] \; \ldots \; x_i[K-1]\right)$, $K=365$, $i=0,\ldots,149$, is detrended by a 4th order polynomial at each measurement station $i$ to form $\widetilde{\x}_i$. The data matrix $\X$ is formed from stacking the detrended data $\wtil{\x}_i$.

We compare the prediction errors of assuming the detrended data $\wtil\x_i$ are generated by the CGP or the MRF models, or by an undirected distance graph as described in~\cite{sandryhaila_discrete_2013}. The distance or geometric graph model uses an adjacency matrix $\A^{\textrm{dist}}$ to model the process\par
{\centering $\x[k]=\w[k]+\sum\limits_{i=1}^{M} h_i (\A^{\textrm{dist}}) \x[k-i]$, \par
}
\noindent
where $h_i (\A^{\textrm{dist}})=\sum\limits_{j=0}^{L_h} c_{ji} (\A^{\textrm{dist}})^j$ are polynomials of order $L_h$ of the distance matrix with elements chosen as \par
{\centering $\displaystyle \A^{\textrm{dist}}_{mn}=\frac{e^{-d_{mn}^2}}{\sqrt{ \sum_{j\in \mathcal{N}^\alpha_n}e^{-d_{nj}^2}\sum_{\ell\in \mathcal{N}^\alpha_m}e^{-d_{m\ell}^2} }}
$ \par}
\noindent
with $\mathcal{N}^\alpha_n$ representing the neighborhood of the $\alpha$ cities nearest to city $n$ and $d_{mn}$ being the geographical Euclidean distance between cities $m$ and $n$. In this model, the number of time lags $M$ is taken to be fixed, and the polynomial coefficients $c_{ji}$ are to be estimated. In our experiments, we assumed $M=2$.

We separated the data into two subsequences, one consisting of the even time indices and one consisting of the odd time indices. One set was used as training data and the other set was left as testing data. This way, the test and training data were both generated from almost the same process but with different daily fluctuations. In this experiment, we compute the prediction MSE as\par
{\centering $\displaystyle MSE=\frac{1}{N(K-M)}\sum\limits_{i=M+1}^{K}\left\| \x[i]-\widehat{\x}[i] \right\|^2$. \par}
Here, since we do not have the ground truth graph for the temperature data, the experiments can be seen as corresponding to the task of prediction. The test error indicates how well the estimated graph and corresponding estimated model can predict the data at the following time instance from past observations. The models are estimated using Algorithms 1, 2, and 3 with $f_1(\x,\y)=\frac{1}{2}\|\x-\y\|_2^2$ and $g_1(\x)=\lambda_1\|\x\|_1$. We repeated the training step for all values of $\lambda_1,\lambda_2,\lambda_3$ on a grid discretizing the interval $(0, 500]$. We used the values of the polynomial coefficient regularization parameter $\lambda_2$ and the commutativity regularization parameter $\lambda_3$ corresponding to the lowest training error to estimate the model on test data and determine the test error; the sparsity regularization parameter $\lambda_1$ directly affects the nonzero proportion and number of nonzero parameters, as expected.

\begin{figure}[t]
	
	%
	\subfigure[Testing Errors vs Sparsity]{\hspace{-0.01\textwidth}%
		\centering
		\centerline{\includegraphics[width=1.01\linewidth]{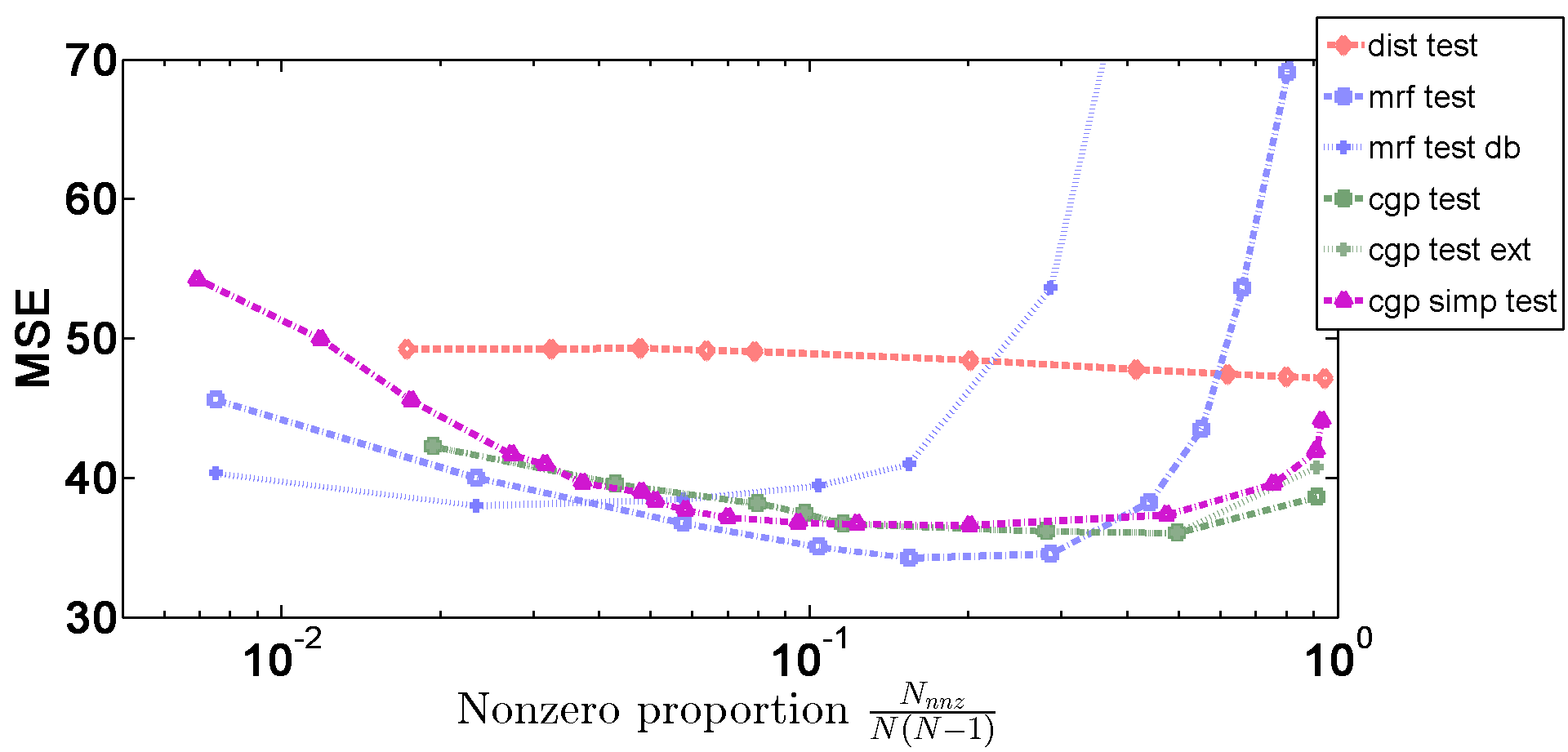}\vspace{-0.25cm}}
		\label{fig:temp_pred_a}
	}
	
	\subfigure[Testing Errors vs Nonzero Parameters]{\hspace{-0.01\textwidth}%
		\centering
		\centerline{\vspace{-0.25cm}\includegraphics[width=1.01\linewidth]{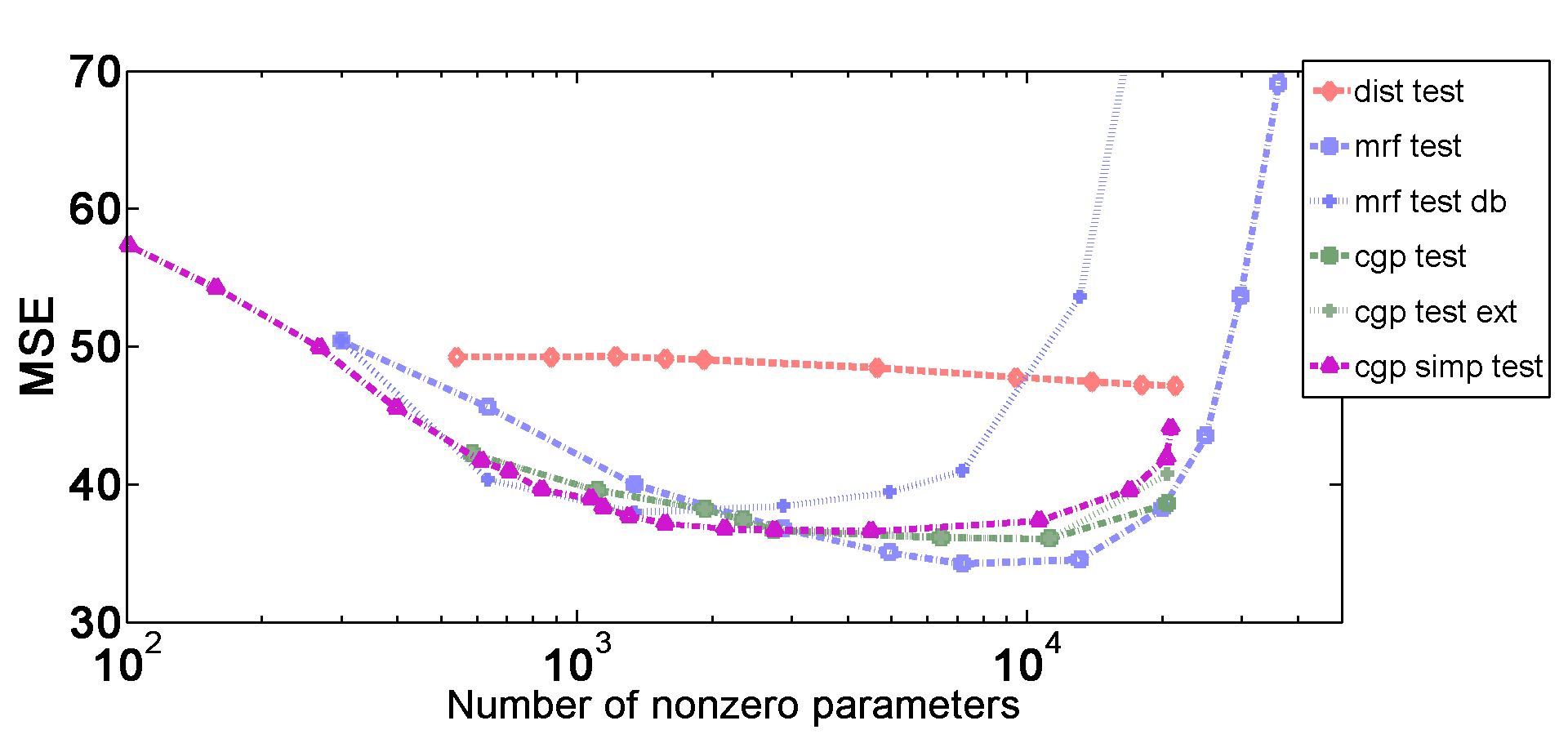}}
		\label{fig:temp_pred_b}
	}	
	\caption{Compression and Prediction Error vs Nonzeros using order $M=2$ model}
	\vspace{-0.5cm}
	\label{fig:temp_pred}
\end{figure}

Figure~\ref{fig:temp_pred_a} shows the performance of the basic, extended, and simple Algorithms~\ref{alg:AlgBasic},~\ref{alg:AlgSimp}, and~\ref{alg:AlgExt}, as well as the distance graph and MRF models, as a function of edge sparsity of the respective graphs. We see that directed graphs estimated from data (either CGP or MRF) perform better in testing than the undirected distance graphs that are derived independently of the data. In addition, for high sparsity or low number of nonzero entries in the estimated $\what\A$, ($p_{\textrm{nnz}}=\frac{N_{\textrm{nnz}}}{N(N-1)}<0.3$, where $p_{\textrm{nnz}}$ is the proportion of nonzero edges in the graph and $N_{\textrm{nnz}}$ is the number of nonzero edges in the graph, not including self-edges), the performance of the CGP is competitive with the MRF model. At lower sparsity levels ($p_{\textrm{nnz}}>0.3$), i.e., denser graphs, the MRF model performs better than the CGP model in minimizing training error (not shown), but not in testing.

Figure~\ref{fig:temp_pred_b} shows the prediction performance of the same algorithms as in~\ref{fig:temp_pred_a} as a function of the total number of nonzero parameters of the respective models. The same trends are present here as in the discussion above. Here, for the CGP model, the number of nonzero parameters is calculated as $N^{CGP}_{\textrm{param}}=N^{CGP}_{\textrm{nnz}}+N+M^{CGP}_{\textrm{nnz}}$, where $N$ includes the diagonal entries and $M^{CGP}_{\textrm{nnz}}\le M(M+1)-3$ counts the nonzeros in $\c$. For the MRF model, the number of nonzero parameters is calculated as $N^{MRF}_{\textrm{param}}=M(N^{MRF}_{\textrm{nnz}}+N)$, where $N$ includes the diagonal entries and the factor of $M$ accounts for the fact that the nonzero entries of $\A^{(i)}$ can be different across $i$. We can see that for the same level of sparsity, the MRF model has more nonzero parameters than the CGP model by approximately a factor of $M$. Here, the CGP has lower test error than MRF with fewer nonzero parameters ($N_{nnz}<2000$).


\begin{figure}[tb]
	\vspace{-1cm}%
	\hspace{0.01\textwidth}%
	\centering\centerline{\includegraphics[width=0.525\textwidth]{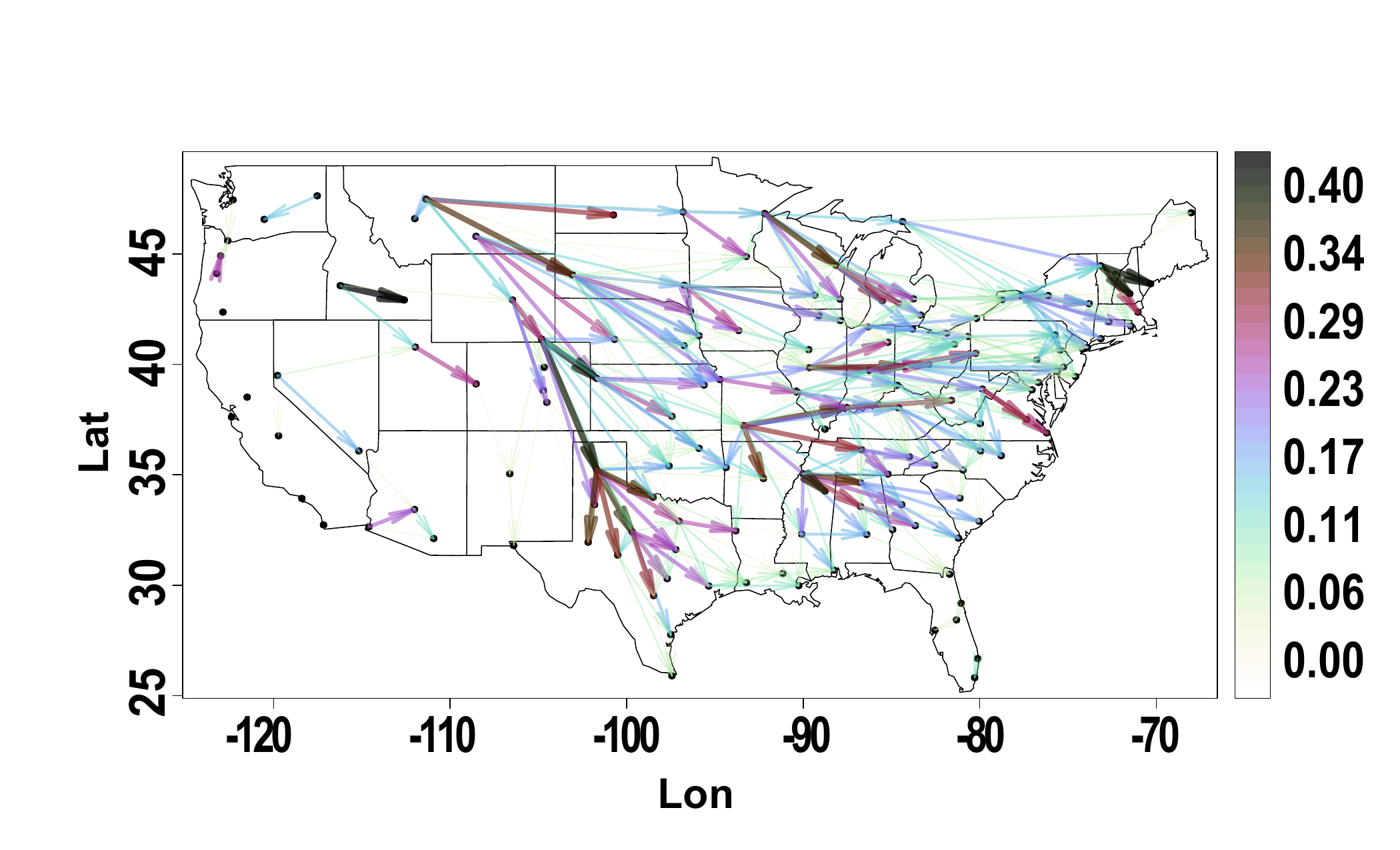}}\vspace{-0.25cm}
	\caption{Estimated CGP temperature graph using order $M=2$ model with sparsity $p_{\textrm{nnz}}=0.05$}
	\label{fig:temp}
\end{figure}

In figure~\ref{fig:temp}, we visualize the temperature network estimated on the entire time series using the CGP model that has sparsity level $p_{\textrm{nnz}}=0.05$.
The $x$-axis corresponds to longitude while the $y$-axis corresponds to latitude.
We see that the CGP model clearly picks out the predominant west-to-east direction of wind in the $x\ge -95$ portion of the country, as single points in this region are seen to predict multiple eastward points.
It also shows the influence of the roughly north-northwest-to-south-southeast Rocky Mountain chain at $-110\le x\le-100$.
This CGP graph paints a picture of US weather patterns that is consistent with geographic and meteorological features.
This consistency may be the most pleasing and surprising observation of this experiment.
This also helps intuitively explain why the distance graph, which is not derived from data, is not as good at predicting weather trends, since cities that are geometrically close may be geographically separated.

\subsection{Synthetic Data}
\label{subsec:syth}
We test our simplified algorithm with $f_1(\x,\y)=\|\x-\y\|_2^2$ and $g_2(\x)=\|\x\|_1$ on larger synthetic datasets to empirically verify the theory developed in section~\ref{sec:Conv}.

\begin{figure*}[t]
	\subfigure[True $\A$ (left) and Estimated $\what{\A}$ (right) for K-Regular graph]{
		\hspace{-0.05\linewidth}
		\includegraphics[width=.525\linewidth]{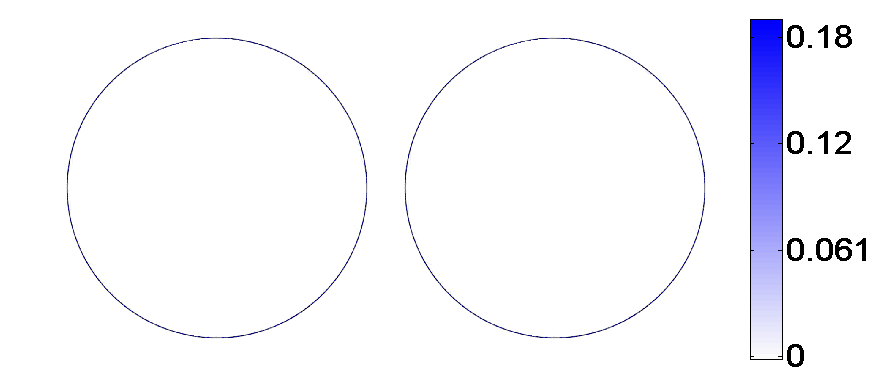}
		\label{fig:toy_top_a}
	}%
	\subfigure[True $\A$ (left) and Estimated $\what{\A}$ (right) for Stochastic Block-Model graph]{
		\hspace{-0.025\linewidth}
		\includegraphics[width=0.525\linewidth]{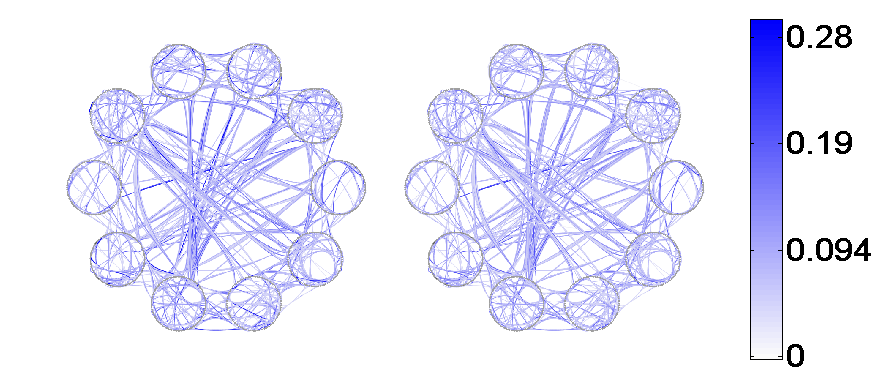}
		\label{fig:toy_top_b}
	}
	\vspace{-0.2cm}
	\subfigure[True $\A$ (left) and Estimated $\what{\A}$ (right) for Erd\"{o}s-Renyi graph]{
		\hspace{-0.05\linewidth}
		\includegraphics[width=0.525\linewidth]{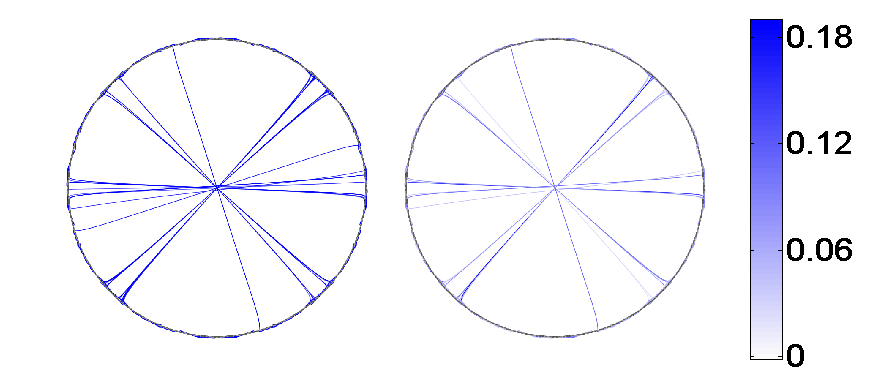}
		\label{fig:toy_top_c}
	}%
	\subfigure[True $\A$ (left) and Estimated $\what{\A}$ (right) for Power Law graph]{
		\hspace{-0.025\linewidth}
		\includegraphics[width=0.525\linewidth]{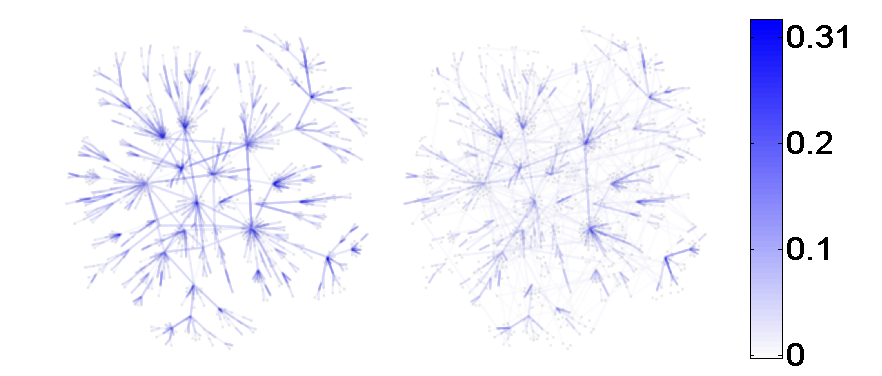}
		\label{fig:toy_top_d}
	}
	\vspace{-0.2cm}
	\caption{True (left) and estimated (right) edge weights (absolute values) for K-Regular (top left), Stochastic Block-Model (top right) Erd\"{o}s-Renyi (bottom left), and Power Law (bottom right) graphs for N=1000\vspace{-0.25cm}}
	\label{fig:tops}
\end{figure*}

The random graphs corresponding to $\A$ were generated with 4 different topologies: K-regular (KR), Stochastic Block-Model (SBM)~\cite{karrer_stochastic_2011}, Erd\"{o}s-Renyi (ER)~\cite{erdos_evolution_1960}, and Power Law (PL).

The KR graph was generated by taking a circle graph and connecting each node to itself using a weight of $-1$ and to its $\xi=3$ neighbors (we use $\xi$ for this quantity rather than K, since we have used $K$ to denote a different quantity) to each side on the circle with weights drawn from a random uniform distribution $\Ucal(0.5,1)$. This resulted in a $(2\nu+1)$-diagonal (in this case heptadiagonal) matrix. Finally the matrix was normalized by 1.5 times its largest eigenvalue.

The SBM graph was generated by creating $10$ clusters with each of the 1000 nodes having uniform probability of belonging to a cluster. Edges between nodes were generated according to assigned intra- and inter- cluster probabilities summarized by a $10\times 10$ matrix (intra-cluster probabilities were on the diagonal, and inter-cluster probabilities on the non-diagonal entries). This matrix was generated randomly and sparsely. Starting with $0.05\I$, we added an independent random quantity to each element, where the variables were uniformly distributed on $[0,0.04)$. Then, entries below $0.025$ were thresholded to $0$. The edges generated were assigned weights from a Laplacian distribution with rate $\lambda_e=2$. Finally, the matrix was normalized by 1.1 times its largest eigenvalue.

The ER graph was generated by taking edges from a standard normal $\Ncal(0,1)$ distribution and then thresholding edges to be between 1.6 and 1.8 in absolute value to yield an effective probability of an edge $p_{ER}\approx 0.04$. The edges were soft thresholded by 1.5 to be between 0.1 and 0.3 in magnitude. Finally, the matrix was normalized by 1.5 times its largest eigenvalue.

The PL graph was generated by starting with a 15 node ER graph with connection probability $0.8$. New nodes were connected by two new edges to and from an existing node of weight drawn as $\Ncal(0,1)$ and then offset $0.25$ away from 0. The connections were made according to a modified preferential attachment scheme~\cite{barabasi_emergence_1999} in which the probability of the new node connecting to an existing node was proportional to the existing node's total weighted degree. The diagonal was set to $-1/2$. Lastly, the matrix was normalized by 1.5 times its largest eigenvalue.

Examples of these topologies with their weighted edges and representative estimates $\what{\A}$ can be seen in figure~\ref{fig:tops} (higher absolute weights are displayed in darker blue). Because of the large number of nodes, these topologies are difficult to visualize. So, \emph{for display purposes only}, we chose parameters that lean towards fewer false alarms at the expense of having more missed edges\textemdash since more false alarms obstruct and obscure the layout. For KR, see figure~\ref{fig:toy_top_a}, we used a circular layout in which the true edges are along the perimeter and not through the interior. This is replicated in the estimated graph\footnote{In the estimated graph, there are some edges through the interior with much lower weights relative to the weights on the edges along the perimeter, which are not visible when visualized.}. For the KR graph whose results are in figure~\ref{fig:toy_top_a} (false alarm rate ($P_{FA}$) $3\times 10^{-4}$ and miss rate ($P_{M}$) $0.16$), we used a circular layout (as previously mentioned); for SBM, see figure~\ref{fig:toy_top_b} (displayed plot has $P_{FA}=7\times 10^{-3}$ and $P_{M}=0.22$), and ER, see figure~\ref{fig:toy_top_c} (displayed plot with $P_{FA}=4\times 10^{-4}$ and $P_{M}=0.58$), we used force-directed edge bundling~\cite{holten_force-directed_2009} to group nearby edges into few thicker ``strands'' in addition to circular layouts; and for PL, see figure~\ref{fig:toy_top_d} (displayed plot with $P_{FA}=4\times 10^{-3}$ and $P_{M}=0.69$), we used a Fruchterman-Reingold~\cite{fruchterman_graph_1991} node positioning.
If we lower the value of the sparsity regularization parameter $\lambda_1$ on our grid, we can reduce the miss rate, while increasing the false alarm rate. Just for reference, we provide another pair of corresponding $P_{FA}$ and $P_{M}$ values: for KR, $P_{FA}=0.01$ and $P_{M}=0.07$; for SBM, $P_{FA}=0.01$ and $P_{M}=0.40$; for ER, $P_{FA}=0.03$ and $P_{M}=0.25$; for PL, $P_{FA}=0.03$ and $P_{M}=0.52$. Playing with $\lambda_1$ we can get other tradeoffs more suitable to specific applications. Just from these experiments, it seems that the SBM and PL models are more difficult to estimate than the KR and ER models. This could be due to the average degree of SBM being high and the maximum degree of PL being high while the number of time observations $K$ is held constant throughout these experiments, which would violate Assumption~\ref{asu:CGP_sparse} or~\ref{asu:CGP_samples}. A more thorough understanding of these tradeoffs, such as what additional conditions need to be satisfied for relevant guarantees to hold, and what broad classes of graphs satisfy these conditions, is left as a topic of future investigation.

Once the $\A$ matrix was generated with $N\in\{1000,1500\}$ nodes, for fixed $M=3$, the coefficients $c_{ij}$ for $2\le i\le M$ and $0\le j\le i$ were generated sparsely from a mixture of uniform distributions $2^{i+j}c_{ij}\sim\frac{1}{2}\Ucal(-1,-0.45)+\frac{1}{2}\Ucal(0.45,1)$ and normalized by $1.5$ to correspond to a stable system.

\begin{figure*}[tb]
	\subfigure[Error in $\A$ (left) and $\epsilon$ (right) for KR graph]{
		\hspace{-0.025\linewidth}
		\includegraphics[width=0.25125\linewidth]{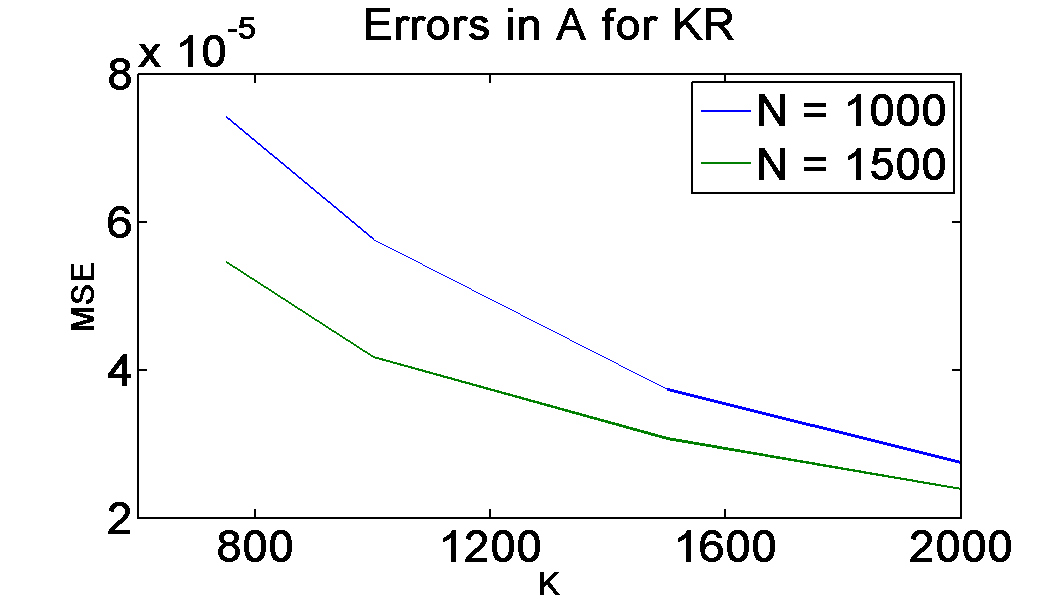}
		\includegraphics[width=0.25125\linewidth]{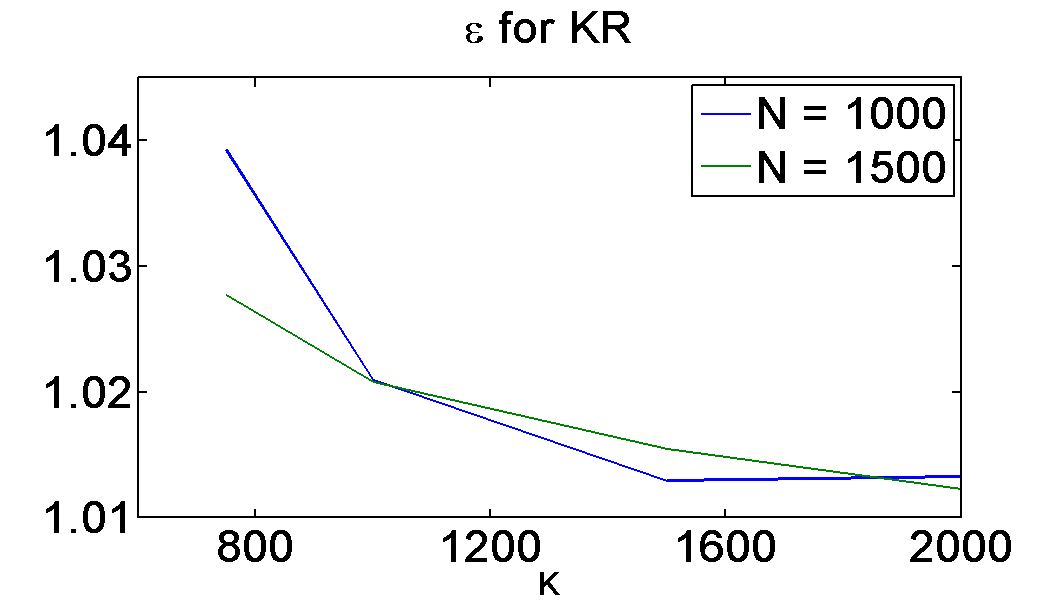}
		\label{fig:toy_err_a}
	}%
	\subfigure[Error in $\A$ (left) and $\epsilon$ (right) for SBM graph]{
		\hspace{-0.0125\linewidth}
		\includegraphics[width=0.25125\linewidth]{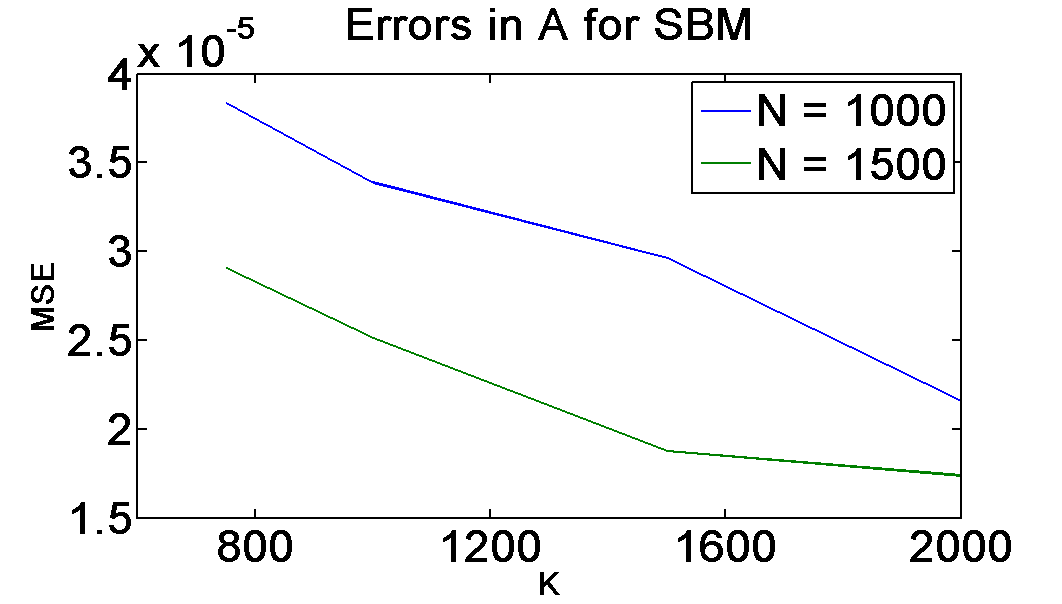}
		\includegraphics[width=0.25125\linewidth]{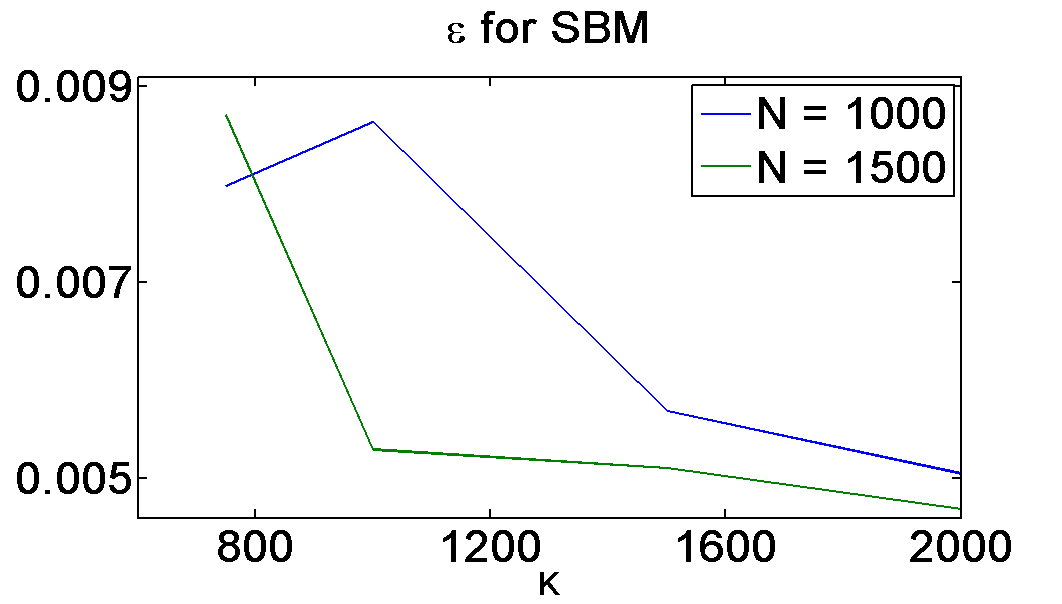}
		\label{fig:toy_err_b}
	}
	
	\subfigure[Error in $\A$ (left) and $\epsilon$ (right) for ER graph]{
		\hspace{-0.025\linewidth}
		\includegraphics[width=0.25125\linewidth]{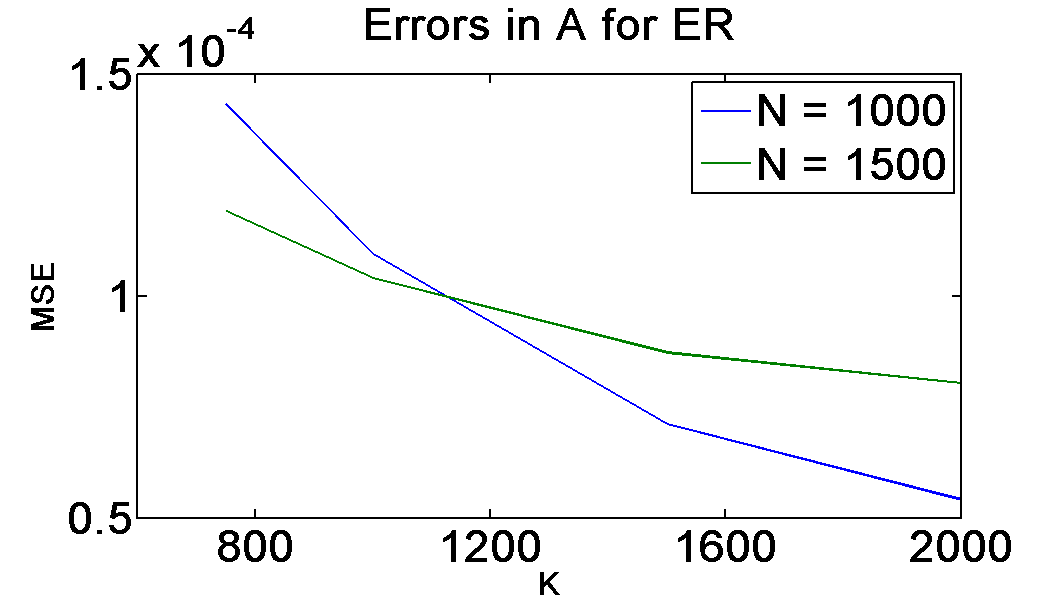}
		\includegraphics[width=0.25125\linewidth]{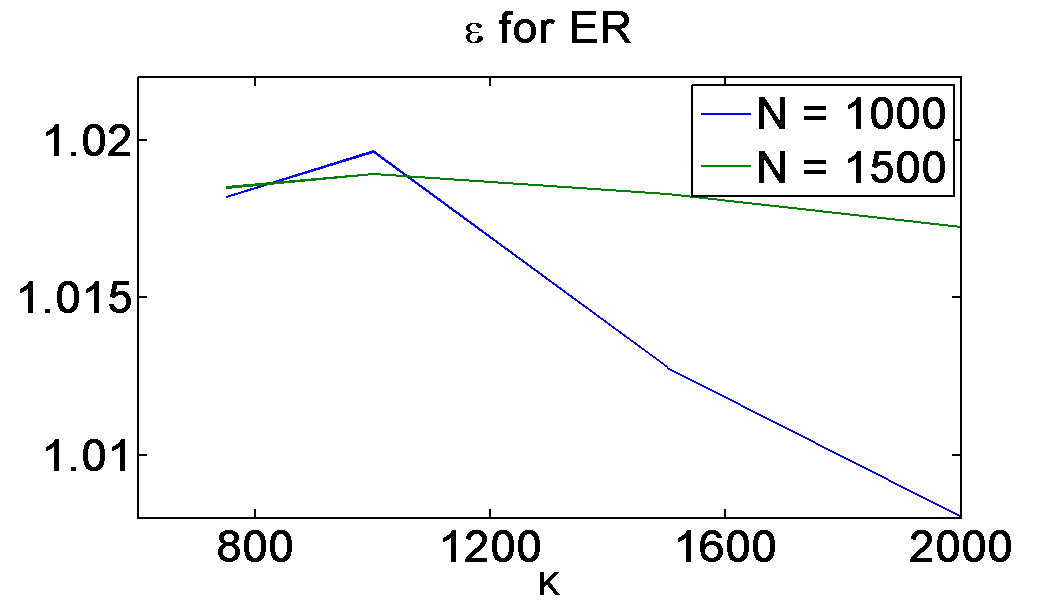}
		\label{fig:toy_err_c}
	}%
	\subfigure[Error in $\A$ (left) and $\epsilon$ (right) for PL graph]{
		\hspace{-0.0125\linewidth}
		\includegraphics[width=0.25125\linewidth]{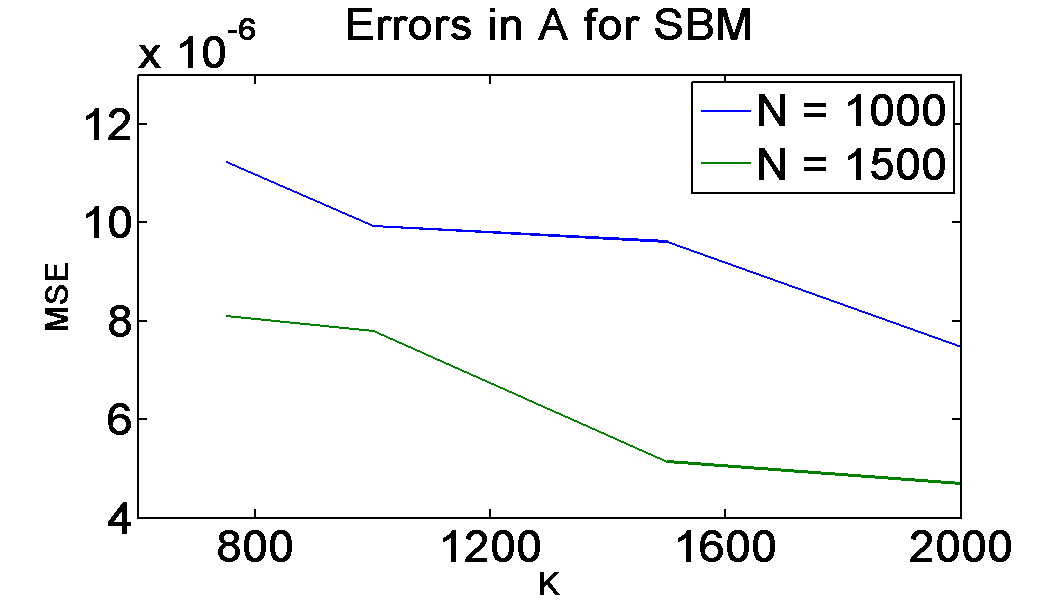}
		\includegraphics[width=0.25125\linewidth]{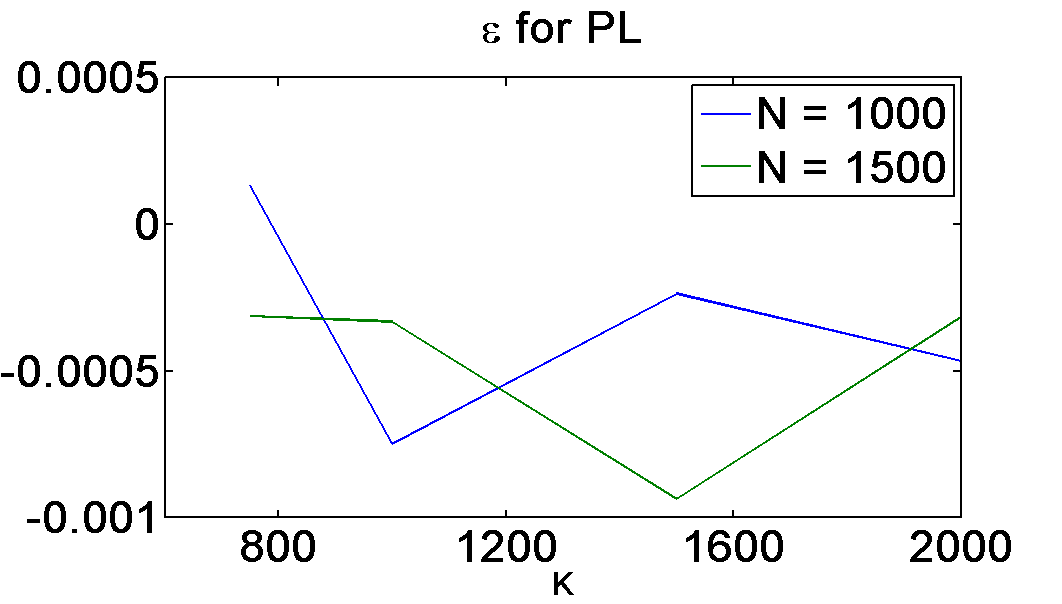}
		\label{fig:toy_err_d}
	}
	\caption{Error in $\A$ (left) and $\epsilon$ (right) for K-Regular (top left), Stochastic Block-Model (top right), Erd\"{o}s-Renyi (bottom left), and Power Law (bottom right) graphs for N=1000,1500\vspace{-0.25cm}}
	\label{fig:sims}
\end{figure*}

The data matrices $\X$ were formed by generating random initial samples (with $500$ burn in samples to reach steady state) and zero-mean unit-covariance additive white Gaussian noise $\w[k]$ computing $K\in\{750,1000,1500,2000\}$ samples of $\x[k]$ according to~(\ref{eq:CGPmodel}). Then 20 such Monte-Carlo data matrices were generated independently, and $(\what{\A},\what{\c})$ was estimated using the simplified algorithm for varying values of $\lambda_1$ and $\rho$. The average error of the $\A$ matrix was computed as \par
{\centering $\displaystyle MSE=\frac{1}{N^2}\|\A-\widehat{\A}\|_F^2$.\par}
The empirical value of the error metric $\what{\epsilon}$ from~\eqref{eq:CGP_error_metric} was also measured by generating another 20 independent sets of time series $\z_k$ (as in~\eqref{eq:zk}) at steady state and using the estimates to predict $\x[k]$ using $\X_{k-1}'$.

In figure~\ref{fig:sims}, we show the errors in $\A$ and the empirical estimates of $\epsilon$  for different values of $N$ and $K$. We performed the estimation across a grid of $\lambda_3$ and $\rho$ and choose the lowest observed error to be $\what{\epsilon}$. We see several different behaviors. We observe that the average errors in $\A$ and empirical averages for $\epsilon$ for the KR graph decrease with both larger $N$ and $K$. We observe similar behavior for the SBM graph. This suggests that graphs generated according to these topologies might satisfy the assumptions in section~\ref{sec:Conv}. On the other hand, the error in $\A$ for ER does not display a clear trend in $N$, although observing more samples still improves the estimate of $\A$ and the model prediction performance as expected. For the PL model, the error in estimating $\A$ decreased with increasing $N$ and $K$, but the error $\epsilon$ fluctuated around $0$ with no clear trend.

This varied behavior arises because the different structured and random graph topologies examined tend to exhibit certain network properties that do not all correspond directly to the assumptions in section~\ref{sec:Conv}. In particular, the K-regular graph by its construction does satisfy the assumptions. This is because taking the $n$-th power of the adjacency matrix of a $\xi$-regular graph of this form results in an $(n\xi)$-regular graph, which satisfies the sparsity~\ref{asu:CGP_sparse} with some constants that do not grow too fast with $N$ and $K$. Thus, the results of the KR graph empirically conform to the predictions given by the theory in section~\ref{sec:Conv}. With the other topologies, the behavior of the sparsity constants is not as immediately clear, but the observed results suggest that some of the assumptions could be slightly loosened, or that other network statistics (e.g., diameter or maximum degree) could play a role in the performance.

\section{Conclusions}
\label{sec:Concl}

This paper presents a methodology to estimate the network structure (graph) capturing spatial (inter) and time (intra) dependencies among multiple time series. These data may arise in many different contexts. The data time dependencies are modeled by an auto-regressive (AR) process. The spatial dependencies are captured by describing the matrix coefficients of the AR process as graph polynomial filters~\cite{sandryhaila_discrete_2014}. The paper presents three algorithms to estimate the graph adjacency matrix and parameters of the graph polynomial filters. These algorithms optimize cost functions that combine a model following functional, e.g., an $\ell_2$ error metric, with sparsity regularizers, e.g., $\ell_1$ metric. The paper carries out the convergence and performance analysis of these algorithms under a set of appropriate assumptions. Finally, the paper illustrates the performance of these algorithms with a set of real data (temperature data collected by 150 weather stations covering the US) and with simulated data for four different types of networks. These experiments show the advantages and limitations of the approach.

\bibliographystyle{IEEEbib}
\bibliography{bibl2}
\setlength{\abovedisplayskip}{2pt}
\setlength{\belowdisplayskip}{2pt}
\setlength{\abovedisplayshortskip}{2pt}
\setlength{\belowdisplayshortskip}{2pt}


\begin{IEEEbiography}[{\includegraphics[width=1in,height=1.25in,clip,keepaspectratio]{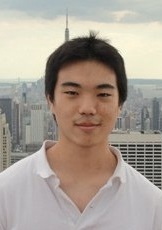}}]{Jonathan Mei}
	received his B.S. and M.Eng. degrees in electrical engineering both from the Massachusetts Institute of Technology in Cambridge, MA in 2013. He is currently pursuing his Ph.D. degree in electrical and computer engineering at Carnegie Mellon University in Pittsburgh, PA. His research interests include computational photography, image processing, reinforcement learning, graph signal processing, non-stationary time series analysis, and high-dimensional optimization. 
\end{IEEEbiography}

\begin{IEEEbiography}[{\includegraphics[width=1in,height=1.25in,clip,keepaspectratio]{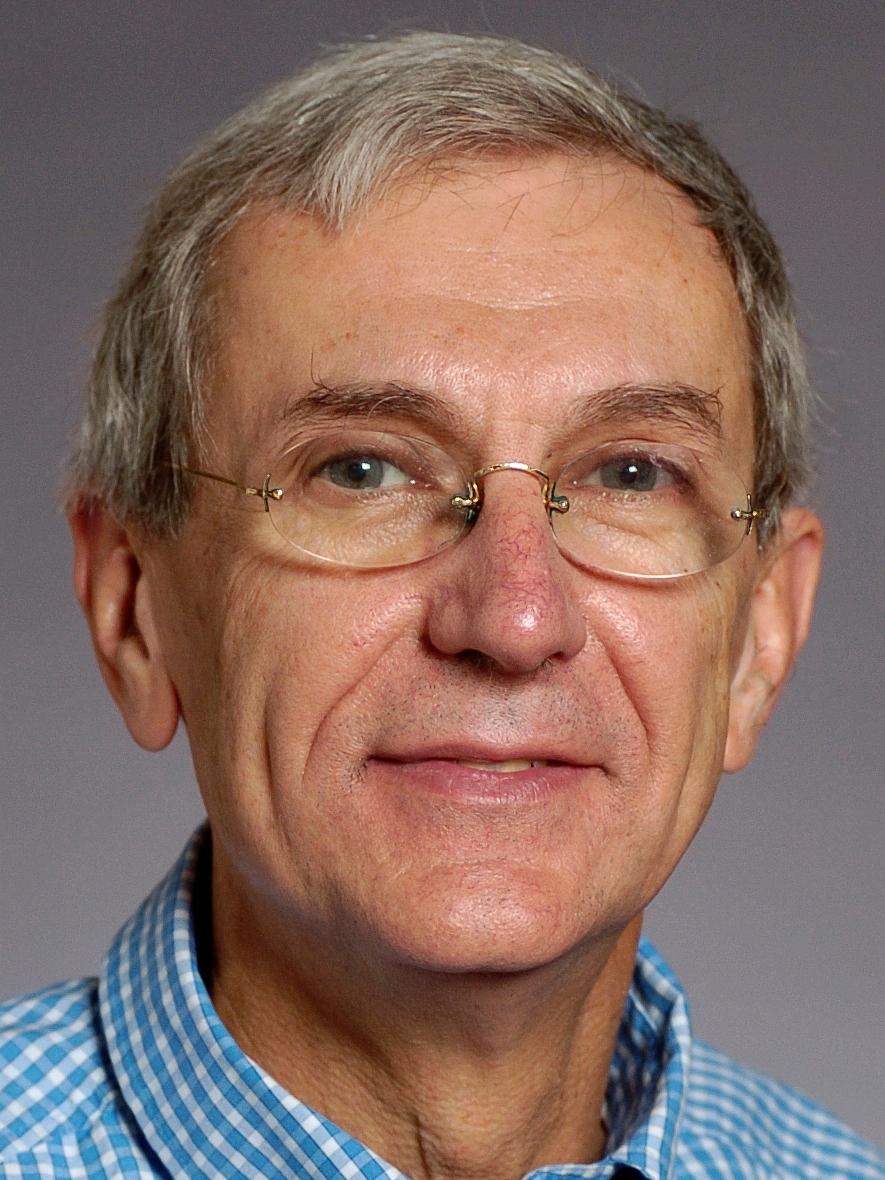}}]{Jos\'e M.~F.~Moura}(S'71--M'75--SM'90--F'94) received the engenheiro electrot\'{e}cnico degree from Instituto Superior T\'ecnico (IST), Lisbon, Portugal, and the M.Sc., E.E., and D.Sc.~degrees in EECS from the Massachusetts Institute of Technology (MIT), Cambridge, MA.
	
	He is the Philip L.~and Marsha Dowd University Professor at Carnegie Mellon University (CMU). He founded and directs a large education and research program between CMU and Portugal, www.icti.cmu.edu.
	
	His research interests are on  data science, graph signal processing, and statistical and algebraic signal and image processing. The technology of two of his patents (co-inventor A. Kav$\check{\textrm{c}}$i\'c) are in about three billion disk drives read channel chips of 60~\% of all computers sold in the last 13 years worldwide and was, in 2016, the subject of the largest university verdict/settlement in the information technologies area.
	
	Dr. Moura is the IEEE Technical Activities Vice-President (2016) and member of the IEEE Board of Directors. Dr. Moura has received several awards, including  the Technical Achievement Award and the Society Award from the IEEE SPS, and the CMU College of Engineering Distinguished Professor of Engineering Award. He is a Fellow of the IEEE, a Fellow of the American Association for the Advancement of Science (AAAS), a corresponding member of the Academy of Sciences of Portugal, Fellow of the US National Academy of Inventors, and a member of the US National Academy of Engineering.
\end{IEEEbiography} 



\vfill


\appendices
\allowdisplaybreaks

\section{Proof of Lemma 2}
\label{app:pf_CGP_estim1}
\setcounter{thm}{1}
\begin{lem}
	\label{lem:CGP_estim1_app}
	Assume~\ref{asu:CGP_model} that $\x[k]$ is generated according to the CGP model with $\A$ satisfying \ref{asu:CGP_sparse}.
	Suppose~\ref{asu:CGP_samples} that the sample size $K$ is large enough. Let $d_i \!>\! 0$ with $i \!=\! 1,...,3$ be universal constants, and let {\small $g(Q)=d_3 \left( 1+\frac{1+Q^2}{(2-Q)^2}\right)$} and $\ell_{MNK}=\sqrt{(\log M+\log N)/K}$. With $\varepsilon_\A=d_1 \exp\{-d_2 K /\omega^2\}$, $\lambda_1 \!=\! 4 g(Q) \ell_{MNK}$, and $\what\R=(\what\R_1,\ldots,\what\R_M)$ the solution to~\eqref{eq:OptRSimpSpecific}, with probability at least $1-\varepsilon_\A$, $\what\A=\what\R_1$ satisfies the inequalities\par
	{\centering \small $\begin{aligned}
		\|\what{\A}-\A\|_1 &\le 256 S_{MN} \ell_{MNK} Q^2 g(Q)  \sigma_u/\sigma_\ell \\
		\|\what{\A}-\A\|_2 \le \|\what{\A}-\A\|_F &\le \delta_{\A}\overset{\Delta}{=} 64 \sqrt{S_{MN}} \ell_{MNK} Q^2 g(Q) \sigma_u/\sigma_\ell\end{aligned}$\par
	}
	\noindent
	where $\omega=\frac{\sigma_u}{\sigma_\ell }\frac{\ Q^2}{\mu_{\min}(\wtil{\Acal})}$ as defined in~\ref{asu:CGP_samples}.			
\end{lem}

First we define several quantities that correspond to the ``stability'' of the system and are used throughout the proof. Let
$\Acal(z)= \I-\sum\limits_{i=1}^{M}P_i(\A,\c)z^i$ and $\wtil\Acal(z)=\I-z\wtil\A$ where $\wtil\A$ is the system matrix for the stacked state in companion form. That is,
\begin{equation}
\label{eq:Atilde}
\wtil{\A}\!=\!\!\left(\!\begin{array}{ccccc}
\A & \!P_2(\A,\c)\! & \!\ldots\! & \P_{M\!-\!1}(\A,\c)\! 	& \!P_M(\A,\c) \\
\I & \0 		& \!\ldots\! & \0 				& \0\\
\0 & \I 		& \!\ldots\! & \0 				& \0 \\
\vdots & \vdots & \!\ddots\! & \vdots 			& \vdots \\
\0 & \0 		& \!\ldots\! & \I 				& \0
\end{array}\!\right)\!,\!
\end{equation}
$\wtil\x[k]=\wtil\A\wtil\x[k-1]+\wtil\w[k]$ where $\wtil\x[k]=(\x[k]^\top \; \ldots \; \x[k-M]^\top)^\top$ and $\wtil\w[k]=(\w[k]^\top \; \0^\top \; \ldots \; \0^\top)^\top$. Then define
\begin{align*}
&\mu_{\max}(\Acal)=\max_{|z|=1}\| \Acal(z)\|_2 \\
&\mu_{\min}(\Acal)=\min_{|z|=1}\| (\Acal(z))^{-1}\|_2^{-1}.
\end{align*}
Now we proceed with the proof.

\begin{proof}[Proof of Lemma 2]	
	To use Propositions 4.1-4.3 from~\cite{basu_regularized_2015}, we bound the quantities $\mu_{\min}(\Acal)$ and $\mu_{\max}(\Acal)$. Letting $\Bcal(z)=\I-\Acal(z)$,
	{\small
		\begin{align}
		\label{eq:B_norm_bnd}
		\sqrt{\mu_{\max}(\Acal)}&\le 1 \!+\! \max_{|z|=1} |z| \left\|\A\right\|_2 \!+\! \sum\limits_{i=2}^{M} \max_{|z|=1} |z^i| \left\|P_i(\A,\c)\right\|_2 \nonumber\\
		&\le 1+L+\sum\limits_{i=2}^{M}\sum\limits_{j=0}^{i} |c_{ij}|\;\|\A^j\|_2 \nonumber\\
		&\le 1+L+L\sum\limits_{i=2}^{M}\sum\limits_{j=1}^{i} |c_{ij}| +\sum\limits_{i=2}^{M}|c_{i0}| \nonumber\\
		&\le (1+L+L\rho+\rho)=Q
		\end{align}
	}by assumption~\ref{asu:CGP_sparse}, and similarly
	\begin{align*}
	\sqrt{\mu_{\min}(\Acal)}&= \min_{|z|=1} \left\|\left(\I-\Bcal(z)\right)^{-1}\right\|_2^{-1}\\
	&\overset{(a)}{\ge} 1- \max_{|z|=1} \left\|\left(\Bcal(z)\right)\right\|_2\ge 2-Q,
	\end{align*}
	where the inequality marked $(a)$ is due to assumption~\ref{asu:CGP_sparse} and the rest follows from similar logic as~\eqref{eq:B_norm_bnd}.
	Propositions 4.2-4.3 hold with high probability when $T\ge d_0 S_{MN}\wtil{\omega}^2 (\log M+\log N)$, where
	\begin{equation}
	\label{eq:omega}
	\omega\ge\wtil{\omega}=\frac{\|\boldsymbol{\Sigma}_\w\|}{\left\|\boldsymbol{\Sigma}_\w^{-1}\right\|^{-1}}\frac{\mu_{\max}(\Acal)}{\mu_{\min}(\wtil{\Acal})}.
	\end{equation}
	Thus, we take $\omega=\wtil{\omega}$, and when assumption~\ref{asu:CGP_samples} holds, this condition is satisfied as well.
	Finally, we substitute these upper and lower bounds for $\mu_{\min}(\Acal)$ and $\mu_{\max}(\Acal)$ into the statements of Proposition 4.1 to yield the statement of Lemma~\ref{lem:CGP_estim1}.
\end{proof}

\section{Proof of Lemma 3}
\label{app:CGP3}
\begin{lem}
	\label{lem:CGP_c}
	Suppose that assumptions~\ref{asu:CGP_model}-\ref{asu:CGP_BA_min} hold. Then for any $0<\beta<\nu <1/2$, and $s_M\le d_4 K/\log n $ and $\lambda_2 \ge q_1$, the solution to~\eqref{eq:FindC2Spec} satisfies
	\[
	P\left( \|\c-\what\c\|_1 \ge \delta_\c\right)\le \varepsilon_\c
	\]
	where
	\[
	\delta_\c=64s_M \lambda_2/\alpha_1
	\]
	and
	\begin{align*}
	\varepsilon_\c=&\varepsilon_{Re}+\varepsilon_\A+\exp\{-d_4 K^{1-2\beta} \}\\
	&+\varepsilon_{De}+(6M+1)\exp\{-d_4 T^{1-2\beta}\},
	\end{align*}
	where $\varepsilon_{Re}$ and $\varepsilon_{De}$ are defined in Propositions~\ref{prop:Re} and~\ref{prop:De} found in subsections~C-D, respectively, $\varepsilon_\A$ is defined in Lemma~\ref{lem:CGP_estim1}, and $\alpha_1$ and $q_1$ are given as
	\begin{equation*}
	\begin{aligned}
	\alpha_1&= \kappa_\B NT/(2K^\nu) + \delta_{Re}K(\tr(\boldsymbol{\Sigma}_0)+G\sqrt{N}) \\
	q_1&= 2 LT^{1-\beta}\sqrt{Nt_N} \pi \sigma_u g(Q)+\delta_\A\what L_M(\delta_\A \!)\sqrt{N} u,
	\end{aligned}
	\end{equation*}
	where
	\begin{equation*}
	\begin{aligned}
	u=& 2 T^{1-\beta}\!\sqrt{t_N} \pi \sigma_u g(Q) \\
	&+ (L \!+\! \delta_\A \what L(\delta_\A)) M \rho K^{1-\beta} \! (\tr(\boldsymbol{\Sigma}_0)+G\sqrt{N})
	\end{aligned}
	\end{equation*}
	and $\delta_{Re}= \delta_\A\what L_M(\delta_\A) n \left(2(1+L)+\delta_\A\what L_M(\delta_\A)\right)$.
\end{lem}

We see that the interpretation of this lemma is similar to that of Lemma~\ref{lem:CGP_estim1}, so that taking $\lambda_2=q_1$ is sufficient to achieve the performance in the lemma. We know that $\delta_\A=\Theta(\sqrt{\log N/K})$. Then with this choice of $\lambda_2$, we have $\alpha_1=\Theta( NK^{1-\nu}+G\sqrt{N}K\delta_{Re})=\Theta(NK+G\sqrt{N}K\delta_\A)=\Theta(NK^{1-\nu} +\sqrt{(N\log N) K})=\Theta(NK^{1-\nu})$ and $u=\Theta(\sqrt{N}K^{1-\beta})$ (from the second term) so that $q_1=\Theta(\sqrt{N}K^{1-\beta}+\sqrt{(N\log N) /K}u)=\Theta(N\sqrt{\log N}K^{1-\beta})$; thus $\delta_\c =O(q_1/\alpha_1)=O\left(\sqrt{\log N/K^{2(\nu-\beta)}}\right)$.
Also, note that
\begin{align*}
\varepsilon_\c \sim& 2\exp\{ -O(K^{1-2\nu})\} + d_1 \exp\{-O(K)\} \\
&+(12M+2)\exp\{-O(T^{1-2\beta})\}\\
\sim& 2\exp\{ -O(K^{1-2\nu})\} + (12M+2)\exp\{-O(T^{1-2\beta})\}\\
\sim& 2\exp\{-O(K^{1-2\nu})\},
\end{align*}
since $K^{1-2\nu}$ is the slowest growing exponent.

To prove this lemma, we need two intermediate results showing that 1) a restricted eigenvalue (Re($\alpha,\tau$)) condition holds with high probability; and 2) a deviation (De($q$)) condition holds with high probability. These two conditions, which will be described shortly, describe the geometric properties of the objective function in the sparse optimization problem. Together, these conditions imply the desired result. These conditions are commonly encountered in sparse estimation, and the proof follows closely~\cite{basu_regularized_2015}.
\subsection{Eigenvalue and Deviation Conditions}
We begin with the statements of the two conditions. Let $n=(M-1)(M+4)/2$, the length of $\c$.

The Re($\alpha,\tau$) condition is satisfied for $\Gamma\in\Rbb^{n\times n}$ if for all vectors $\theta\in \Rbb^{n}$
\begin{equation}
\label{eq:Re_cond}
\theta^\top \Gamma \theta \ge \alpha\|\theta\|_2^2-\tau\|\theta\|_1^2.
\end{equation}
In sparse estimation, this condition is used to show curvature of the objective function is large enough near the true value of the parameter in sparse directions~\cite{loh_high-dimensional_2012}. In other words, large perturbations in the objective function correspond to small perturbations in the parameter estimate.

The De($q$) condition is satisfied for $(\Gamma,\gamma)\in \Rbb^{n\times n}\times\Rbb^{n}$ if for the true value of the parameter $\c$,
\begin{equation}
\label{eq:De_cond}
\|\gamma -\Gamma\c \|_\infty \le q.
\end{equation}
In sparse estimation, this condition intuitively states that the gradients of objective function are small near the true parameter value~\cite{loh_high-dimensional_2012}. In other words, the estimated parameter value that minimizes the objective function is near the true parameter value.

\subsection{Discussion}
For our problem, we wish to show that Re($\alpha_1,\tau$) and De($q_1$) are satisfied for $(\what\Gamma,\what\gamma)=\left(\B(\what\A)^\top \B(\what\A),\B(\what\A)^\top\Y(\what\A)\right)$ with high probability. If these two conditions hold for these matrices, then the estimate $\what\c$ will satisfy $\|\what\c-\c\|_1 \le 64 s_{M}\lambda_2 /\alpha_1$ for $\lambda_2 \!\ge\! q_1$.

Usually, in performance analysis of sparse estimation, these two restricted eigenvalue and deviation conditions would be shown using the true data (i.e., in our problem, that Re($\alpha_0,\tau$) and De($q_0$) are satisfied for $(\Gamma,\gamma)=\left(\B(\A)^\top \B(\A),\B(\A)^\top\Y(\A)\right)$). Matrices $\B(\A)$ and $\Y(\A)$ have entries that can be described by a multivariate Gaussian variable, so showing the two conditions on $(\Gamma,\gamma)$ is relatively straightforward.
However in our two-stage estimation, we can only use the estimated parameter $\what\A$ instead of the true parameter $\A$. Thus, the challenge in showing that these two conditions are satisfied comes from only being able to use the matrices $\B(\what\A)$ and $\Y(\what\A)$ that are complicated random functions of the true parameters and data, and are no longer described by a multivariate Gaussian variable.

Our approach is to show that if the two conditions are satisfied on the matrices $\B(\A)$ and $\Y(\A)$, then the two conditions will also be satisfied on the matrices $\B(\what\A)$ and $\Y(\what\A)$ that are actually available at the second stage. Thus, we rely on two additional intermediate results, which we prove first, showing that these two conditions are indeed satisfied on the matrices $\B(\A)$ and $\Y(\A)$ with high probability.
We additionally point out that although the problem of estimating $\c$ is overdetermined, it may still be ill-conditioned due to the possibly highly correlated form of the data matrices $\B(\A)$ and $\Y(\A)$. Since it is not immediately obvious that the performance would be good, it is important to still demonstrate that these conditions hold.

To summarize the above discussion, for our problem we first show that Re($\alpha_0,\tau$) and De($q_0$) are satisfied for $(\Gamma,\gamma)=\left(\B(\A)^\top \B(\A),\B(\A)^\top\Y(\A)\right)$, and then show that this implies that Re($\alpha_1,\tau$) and De($q_1$) are satisfied for $(\what\Gamma,\what\gamma)=\left(\B(\what\A)^\top \B(\what\A),\B(\what\A)^\top\Y(\what\A)\right)$. We show Re($\alpha_1$,$\tau$) in Proposition~\ref{prop:Re} and De($q_1$) in Proposition~\ref{prop:De}.

\subsection{Proof of Supplementary Proposition 1}
\label{app:CGPS1}
\begin{propx}
	\label{prop:Re}
	Suppose that assumptions~\ref{asu:CGP_model}-\ref{asu:CGP_BA_min} hold. Then for any $\theta\in\Rbb^{n}$ and any $0<\nu<1/2$,
	\[
	P\left(\|\B(\A)\theta\|_2^2 \le \alpha_0\|\theta\|_2^2-\tau\|\theta\|_1^2 \right) \le \varepsilon_{Re}
	\]
	where $\tau=d_5 \alpha_0 (1+L)^4 n^2 G^2 K^\nu/(2 (\log n) \kappa_\B^2 N T^2)$, $\alpha_0=\kappa_\B NT/2$, and $\varepsilon_{Re}=2\exp\{ -d_4 \kappa_\B ^2 T/((1+L)^4 n^2 G^2 K^{2\nu})\}$.\\
	That is,
	Re$(\alpha_0,\tau)$ holds for estimating $\c$ using $\B(\A)^\top \B(\A)$ with probability at least $1-\varepsilon_{Re}$.
\end{propx}

\begin{proof}
	This proof follows loosely the proof of Proposition 4.2 in~\cite{basu_regularized_2015}, first bounding the quadratic form of $\B(\A)^\top \B(\A)$ using the Hanson-Wright concentration inequality followed by a discretization argument.
	
	Let {\small $\Dcal(\A,\theta)\!=\! \left(\!\begin{array}{cccc} \! \D_0(\A,\theta)^\top \!&\! \D_1(\A,\theta)^\top \!\!&\! ... \!&\! \D_{K\!-\!M\!-\!1}(\A,\theta)^\top \! \end{array}\!\right)^\top$}
	where the block rows of $\Dcal(\A,\theta)$ are
	
	{\small $\D_i(\A,\theta)\!=\!\left(\!\begin{array}{ccccc}\! \0_{N\times Ni} \!&\! P_2(\A,\theta) \!\!&\!\! ... \!\!&\! P_M(\A,\theta) \!&\! \0_{N\times N(K\!-\!M\!-\!2\!-\!i)} \!\end{array} \!\right)$.
	}Then consider any vector $\|\theta\|\le 1$,
	{\small
		\begin{equation}
		\begin{aligned}
		\label{eq:app_BA_theta_norm}
		\|\B(\A)\theta\|_2^2 &= \sum\limits_{k=M}^{K-1}\left\|\sum\limits_{i=2}^M  P_i(\A,\theta)\x[k-i]\right\|_2^2\! \\
		&=\|\Dcal(\A,\theta) \x[0\!:\!K-3]\|_2^2,
		\end{aligned}
		\end{equation}
	}where $\x[0\!:\!K-3]=\left( \begin{array}{ccc}\x[0]^\top & ... & \x[K-3]^\top \end{array}\right)^\top$.
	By an extension of Gershgorin's Circle Theorem to block matrices~\cite{feingold_block_1962},
	\begin{align*}
	\|\Dcal(\A,\theta)\|_2 &\le \sum\limits_{i=2}^{M} \left\| P_i(\A,\theta) \right\|_2 \le \sum\limits_{i=2}^{M} \left\| \sum\limits_{j=0}^{i}\theta_{ij}\A^j \right\|_2\\
	& \le \sum\limits_{i=2}^{M} \sum\limits_{j=0}^{i} |\theta_{ij}| \|\A^j \|_2 \le (1\!+\!L)\sum\limits_{i=2}^{M} \sum\limits_{j=0}^{i}|\theta_{ij} |\\
	&= (1+L)\|\theta\|_1 \le (1+L)\sqrt{n}\|\theta\|_2\\
	&\le (1+L)\sqrt{n}.
	\end{align*}
	Now,
	\begin{flalign}
	\label{eq:app_BA_theta}
	&\|\B(\A)\theta\|_2^2 -\Ebb\left[ \|\B(\A)\theta\|_2^2\right]& \nonumber \\
	&\; \!=\! \|\Dcal(\A,\theta)\x[0\!:\!K\!-\!3]\|_2^2 \!-\! \Ebb\left[ \|\Dcal(\A,\theta)\x[0\!:\!K\!-\!3]\|_2^2 \right]& \nonumber \\
	&\; \overset{(a)}{\le} \|\Dcal(\A,\theta)\|_2^2\|\boldsymbol{\Sigma}_0\|NK\eta\le (1+L)^2 n GNK \eta &
	\end{flalign}
	with probability at least $1-2\exp\{-d_4 NK\, \min(\eta,\eta^2) \}$ for some universal constant $d_4$, where $(a)$ follows from the Hanson-Wright inequality~\cite{rudelson_hanson-wright_2013}. Then from the discretization argument of Lemma F.2~\cite{basu_regularized_2015}, for an integer $s\ge 1$ (to be specified later) and set $\Kcal_{2s}=\{\theta\in\Rbb^{n} \, : \, \|\theta\|\le 1, \|\theta\|_0 \le 2s \}$,
	\begin{equation*}
	\begin{aligned}
	P&\!\left(\!\sup\limits_{\theta \in \Kcal_{2s}}\|\B(\A)\theta\|_2^2-\Ebb\left[ \|\B(\A)\theta\|_2^2\right] \!\ge\! (1+L)^2 n NK G \eta\!\right)\!\\ &\le\! 2 \exp\{ \!-\! d_4 NK\min(\eta,\eta^2) \!+\! 2s\min(\log n,\log (21e n/2s ))\}.
	\end{aligned}
	\end{equation*}
	Finally, from Lemmas 12 and 13 in~\cite{loh_high-dimensional_2012} and taking\\
	$s \!=\! \lceil d_4 NK\eta^2/(4\log n) \rceil$ and $\eta \!=\! \kappa_\B NT/[54(1+L)^2 nGNK^{\nu}] \!\le\! 1$ with $0<\nu<1/2$ so that $\eta^2 \le \eta$ where $\kappa_\B NT $ is the minimum singular value of $\Ebb[\B(\A)^\top \B(\A)]$ as defined in~\ref{asu:CGP_BA_min}, we have $\B(\A)^\top\B(\A)$ satisfies Re$(\kappa_\B NT/(2K^\nu),\kappa_\B NT /(2sK^\nu))$ with probability at least $1-2\exp\{ -d_4 \kappa_\B ^2 T/((1+L)^4 n^2 G^2 K^{2\nu})\}$.
\end{proof}
\subsection{Proof of Supplementary Proposition 2}
\label{app:CGPS2}
\begin{propx}
	\label{prop:De}
	Suppose that assumptions~\ref{asu:CGP_model}-\ref{asu:CGP_BA_min} hold. Then for any $0<\beta<1/2$,
	\[
	P(\|\B(\A)^\top \e \|_\infty \ge q_0)\le \varepsilon_{De},
	\]
	where $q_0=2 LT^{1-\beta}\sqrt{Nt_N} \pi \sigma_u g(Q)$ and $\varepsilon_{De}=6M \exp\{-d_4 T^{1-2\beta}\}$.\\
	That is, De$(q_0)$ holds for estimating $\c$ using $\Y(\A)\B(\A)$ with probability at least $1-\varepsilon_{De}$.
\end{propx}

\begin{proof}
	The proof is a straightforward application of Proposition 2.4 from~\cite{basu_regularized_2015} and the union bound (as in the proof of Proposition 4.3). Letting $\e \!=\! \Y(\A) \!-\! \B(\A)\c$ and $T=K-M$,
	{\small
		\begin{equation*}
		\begin{aligned}
		&\left\| \B(\A)^\top \e \right\|_\infty \!\!\!= \! \max\limits_{{\tiny\begin{array}{c}1 \!\le\! i \!\le\! M \\ 1 \!\le\! j \!\le\! i \end{array}}}\left|\left(\sum\limits_{k \!=\! M}^{K \!-\! 1}\! \x[k-i]^\top (\A^j)^\top\w[k] \right)\right| \\
		&\: \le \!\!{\small \max\limits_{{\tiny\begin{array}{c}1 \!\!\le\!\! i \!\!\le\!\! M \\ 1 \!\le\! j \!\le\! i \end{array}}}\!\!\!\!\left|\tr\left(\!( \A^j)^\top \! \w[M \!:\! K \!-\! 1 ]\x[M \!-\! i \!:\! K \!-\! 1 \!-\! i]^\top \!\right)\! \right|}\\
		&\: \le \!\!\!{\small \max\limits_{{\tiny\begin{array}{c}1 \!\le\! i \!\le\! M \\ 1 \!\le\! j \!\le\! i \end{array}}} \!\!\! T\|\A^j\|_1  \!\left\|\w[M \!:\! K \!-\! 1 ]\x[M \!-\! i \!:\! K \!-\! 1 \!-\! i]^\top \!/\! T \right\|_{\infty} }\\
		&\: \overset{(a)}{\le} T \max\limits_{{\tiny 1 \!\le\! j \!\le\! M }}\sqrt{t_N}\|\A^j\|_F \left( \pi \sigma_u\left(1 \!+\! \frac{1 \!+\! \mu_{max}(\Acal)}{\mu_{min}(\Acal)}\right)\eta\right)\\
		&\: \overset{(b)}{\le} 2 \sqrt{t_N}T \max\limits_{{\tiny 1 \!\le\! j \!\le\! M }}\sqrt{N}\|\A^j\|_2 \left(\pi \sigma_u g(Q) \eta\right) \\
		&\: \le 2 LT\sqrt{Nt_N} \pi \sigma_u g(Q)\eta
		\end{aligned}
		\end{equation*}
	}with probability at least $1-6M \exp\{-d_4 T\,\min(\eta,\eta^2)\}$, where $t_N$ is the maximum sparsity of $\A^j$ as defined in assumption~\ref{asu:CGP_sparse}, $(a)$ is implied by Proposition 2.4 from~\cite{basu_regularized_2015} and $(b)$ is implied by the analysis in Lemma~\ref{lem:CGP_estim1_app}. To finish the proof, we take $\eta=T^{-\beta}$ for any $0<\beta<1/2$, noting that for this choice, $\eta^2 < \eta$.
\end{proof}

\subsection{Proof of Lemma 3}
\label{app:CGP3_rep}
Armed with these two results, we resume with our lemma.
\begin{proof}
	
	From Lemma~\ref{lem:CGP_estim1}, we have that $P(\|\what\A \!-\! \A\|_2 \!\ge\! \delta_\A) \!\le\! \varepsilon_\A$.
	First, we consider the Re condition. Let $\Delta_\alpha=\alpha_1-\alpha_0$ and $\wtil\Bcal=\B(\what\A)^\top\B(\what\A)-\B(\A)^\top \B(\A)$. Then,
	{\small \begin{align}
		&P\left(\|\B(\what\A)\theta\|_2^2 \le \alpha_1\|\theta\|_2^2-\tau\|\theta\|_1^2 \right) \nonumber\\
		&=\! P\left(\|\B(\what\A)\theta\|^2 \le (\alpha_0 \!+\! \Delta_\alpha)\|\theta\|^2 \!-\! \tau\|\theta\|_1^2 \bigcap \|\wtil\Bcal\|_2 \!<\! \Delta_\alpha \right)\nonumber\\*
		&\quad + P\left(\|\B(\what\A)\theta\|_2^2 \le \alpha_1\|\theta\|_2^2-\tau\|\theta\|_1^2 \bigcap \|\wtil\Bcal\|_2 \ge \Delta_\alpha \right) \nonumber\\
		&\le P\left( \|\B(\what\A) \theta\|_2^2 \!\le\! \alpha_0\|\theta\|_2^2 \!+\! \theta^\top\wtil\Bcal\theta \!-\! \tau\|\theta\|_1^2 \bigcap \|\wtil\Bcal\|_2 \!<\! \Delta_\alpha \right) \\*
		&\quad + P\left(\|\wtil\Bcal\|_2 \ge \Delta_\alpha \right) \nonumber\\
		&\le P\left( \| \B(\A)\theta\|_2^2 \le \alpha_0\|\theta\|_2^2-\tau\|\theta\|_1^2  \right) \nonumber\\*
		&\quad \!+\! P\left(\|\wtil\Bcal \|_2 \!\ge\! \Delta \bigcap \|\what\A-\A\|_2 \!\le\! \delta_\A \right) \!+\! P\left(\|\what\A-\A\|_2 \!\ge\! \delta_\A \right) \nonumber\\
		&\le \varepsilon_{Re}+\varepsilon_\A + P\left(\|\wtil\Bcal \|_2 \ge \Delta_\alpha \bigcap \|\what\A-\A\|_2\le \delta_\A \right). \nonumber
		\end{align}
	}
	We bound the last probability by manipulating the first term,
	{\small
		\begin{equation}
		\begin{aligned}
		\|\wtil\Bcal \|_2&= \|\B(\what\A)^\top \B(\what\A)-\B(\A)^\top \B(\A)\|_2\\
		&\le \|\B(\what\A)^\top \B(\what\A) - \B(\what\A)^\top \B(\A) \|_2 \\
		&\quad + \| \B(\what\A)^\top \B(\A) - \B(\A)^\top \B(\A)\|_2\\
		&\le \|\B(\what\A) \!-\! \B(\A)\|_2 \left(2\|\B(\A)\|_2 \!+\! \|\B(\what\A) \!-\! \B(\A)\|_2 \right).
		\end{aligned}
		\end{equation}
	}
	Under the constraint $\|\what\A-\A\|_2\le\delta_\A$, for any vector $\|\theta\|_2\le 1$ following from the same logic as~\eqref{eq:app_BA_theta},
	{\small
		\begin{equation*}
		\begin{aligned}
		\|(\B(\what\A) & \!-\! \B(\A))\theta\|_2^2 = \| (\Dcal(\what\A,\theta)-\Dcal(\A,\theta))\x[0\!:\!K-3] \|_2^2\\
		&\le \left(\delta_\A \what L_M(\delta_\A) \right)^2 n \|\X\|_F^2
		\end{aligned}
		\end{equation*}
	}
	\begin{flalign}
	\textrm{and} \qquad \|\B(\A)\|_2^2 &\le (1+L)^2 n \|\X\|_F^2. &
	\end{flalign}
	Thus, letting \[
	\delta_{Re}\overset{\Delta}{=} \delta_\A\what L_M(\delta_\A) n \left(2(1+L)+\delta_\A\what L_M(\delta_\A)\right),
	\]
	we have
	{\small
		\begin{align}
		\label{eq:pop}
		\|\wtil\Bcal \|_2&\le \delta_{Re} \|\X\|_F^2\\
		\!\Rightarrow\!	P &\left(\|\wtil\Bcal \|_2 \!\ge\! \Delta_\alpha \!\bigcap\! \|\what\A \!-\! \A\|_2 \!\le\! \delta_\A \! \right) \!\le\! P (\| \X \|_F^2 \!\ge\! {\Delta_\alpha / \delta_{Re}} ) \le \varepsilon_\Bcal, \nonumber
		\end{align}
	}where by the Hanson-Wright inequality~\cite{rudelson_hanson-wright_2013}, taking $\Delta_\alpha = \delta_{Re}K^{1-\beta}(\tr(\boldsymbol{\Sigma}_0)+G\sqrt{N})$, we have $\varepsilon_\Bcal = \exp\{-d_4 K^{1-2\beta} \}$.
	Finally,
	{\small \begin{equation}
		\begin{aligned}
		&P\left(\|\B(\what\A)\theta\|_2^2 \le \alpha_1\|\theta\|_2^2-\tau\|\theta\|_1^2 \right) \le \varepsilon_{Re}+\varepsilon_\A + \varepsilon_{\Bcal}.
		\end{aligned}
		\end{equation}
	}	
	Second, letting $\e \!=\! \Y(\A) \!-\! \B(\A)\c$ and $\what\e \!=\! \Y(\what\A) \!-\! \B(\what\A)\c$ and $\Delta_q=q_1-q_0$, we consider the De condition.
	{\small
		\begin{equation*}
		\begin{aligned}
		P&\left( \left\|\B(\what\A)^\top \what\e \right\|_\infty \! \ge q_1 \right) \le P\left( \left\|\B(\A)^\top \e \right\|_\infty \! \ge q_0 \right) \\
		&\qquad + P\left( \left\|\B(\what\A)^\top \what\e - \B(\A)^\top \e \right\|_\infty \! \ge \Delta_q \right) \\
		&\le \varepsilon_{De} + P\left( \left\|(\B(\what\A)^\top - \B(\A)^\top ) \e \right\|_\infty  \! \ge \Delta_{q_1} \right) \\
		&\quad + P\left( \left\|\B(\what\A)^\top (\what\e - \e) \right\|_\infty  \ge \Delta_{q_2} \right),
		\end{aligned}
		\end{equation*}
	}where $\Delta_{q}=\Delta_{q_1}+\Delta_{q_2}$. We examine the term
	\begin{align}
	\label{eq:gnarly_1}
	&\left\|\left(\B(\what\A) \!-\! \B(\A) \right)^\top \e \right\|_\infty \nonumber\\
	&\: \!=\! \max\limits_{{\tiny\begin{array}{c}1 \!\le\! i \!\le\! M \\ 1 \!\le\! j \!\le\! i \end{array}}}\left|\left(\sum\limits_{k=i}^{T \!-\! 1 \!+\! i}\! \x[k-i]^\top (\what\A^j-\A^j)^\top \w[k] \right)\right| \\
	&\: \!=\!\!\! {\small \max\limits_{{\tiny\begin{array}{c}1 \!\le\! i \!\le\! M \\ 1 \!\le\! j \!\le\! i \end{array}}} \!\!\!\!\left|\tr\left((\what\A^j \!-\! \A^j)^\top \!\w[M \!:\! K \!-\! 1 ]\x[M \!-\! i \!:\! K \!-\! 1 \!-\! i]^\top \right) \right|} \nonumber \\
	&\: \!\le\!\! {\small \delta_\A\what L_M(\delta_\A)\sqrt{\!Nt_N \!} \!\! \max_{{\tiny 1\le i \le M}} \! \left\|\w[M \!:\! K \textrm{-} 1 ]\x[M \textrm{-} i \!:\! K \textrm{-} 1 \textrm{-} i]^\top \right\|_{\infty} } \nonumber\\
	&\: \!\overset{(a)}{\le}\! 2\delta_\A\what L_M(\delta_\A) T^{1-\beta}\sqrt{Nt_N} \pi  \sigma_u g(Q) \nonumber
	\end{align}
	with probability at least $1-6M \exp\{-d_4 T^{1-2\beta}\}$, where $t_N$ is defined in~\ref{asu:CGP_L_rho}, and in $(a)$ we again invoke Proposition 2.4 from~\cite{basu_regularized_2015} similarly as in Proposition~\ref{prop:De}.
	
	To finish, we see that
	\begin{equation*}
	\begin{aligned}
	\what\w[k]-\w[k]=\sum\limits_{m=1}^{M}\sum\limits_{\ell=1}^{m}c_{m\ell} (\what\A^\ell \!-\! \A^\ell )\x[k \!-\! m]
	\end{aligned}
	\end{equation*}
	and that $\|\what\A^j\|_2\le \|\A^j+(\what\A^j-\A^j)\|_2\le L+\delta_\A \what L(\delta_\A)$.\\
	These imply,
	\begin{equation*}
	\begin{aligned}
	&\left\|\B(\what\A)^\top (\what\e - \e) \right\|_\infty \\
	&\: \le\!\! \max\limits_{{\tiny\begin{array}{c}1 \!\le\! i \!\le\! M \\ 1 \!\le\! j \!\le\! i \end{array}}}\!\!\left|\left(\sum\limits_{k=i}^{T \!-\! 1 \!+\! i}\!\! \x[k \!-\! i]^\top (\what\A^j)^\top (\what\w[k] \!-\! \w[k]) \right)\right| \\
	&\overset{(a)}{\le} \!\! (L+\delta_\A \what L(\delta_\A))\delta_\A \what L(\delta_\A) \sqrt{N} \\
	&\quad \times \max\limits_{{\tiny 1 \!\le\! i \!\le\! M }} \sum\limits_{m=1}^{M}\left\|\left(\sum\limits_{k=i}^{T \!-\! 1 \!+\! i}\!\! \|\c_{m\cdot}\| \x[k \!-\! i]^\top \x[k \!-\! m] \right)\right\|_F\\
	&\le \!\! (L+\delta_\A \what L(\delta_\A))\delta_\A \what L(\delta_\A) \sqrt{N} M \rho \|\X\|_F^2\\
	&\overset{(b)}\le \! (L+\delta_\A \what L(\delta_\A))\delta_\A \what L(\delta_\A) \sqrt{N} M \rho K^{1-\beta}(\tr(\boldsymbol{\Sigma}_0)+G\sqrt{N})
	\end{aligned}
	\end{equation*}
	with probability at least $1-\exp\{-d_4 K^{1-2\beta} \}$
	where $(a)$ follows from similar logic as~\eqref{eq:gnarly_1} and $(b)$ from similar logic to~\eqref{eq:pop}.
	
	Finally, we arrive at\par
	{\centering \small $\Delta_q \!=\! \delta_\A\what L_M(\delta_\A \!)\sqrt{N} u,$\par
	}where \\
	{\small $u = 2 T\!\sqrt{t_N} \pi \sigma_u g(Q) \!+\! (L \!+\! \delta_\A \what L(\delta_\A)) M \rho K \! (\tr(\boldsymbol{\Sigma}_0)+G\sqrt{N})$}
	and
	$P\left( \left\|\B(\what\A)^\top \what\e \right\|_\infty \! \ge q_1 \right) \le \varepsilon_{De}+(6M+1)\exp\{-d_4 T\}$
	since $T < K$.
\end{proof}

\section{Proof of Theorem 1}
Finally, with the two lemmas in hand, we return to the proof of the main theorem, which we restate for convenience.
\label{app:CGP1}
\setcounter{thm}{0}
\begin{thm}
	\label{thm:main_app}
	For any $0<\beta<\nu<1/2$, and some universal constant $d_1$, assumptions~\ref{asu:CGP_model}-\ref{asu:CGP_BA_min} are sufficient for the error $\epsilon$ in~\eqref{eq:CGP_error_metric} to satisfy
	{\small
		\begin{equation}
		\label{eq:CGP_theorem_app}
		\begin{aligned}
		\epsilon \le \left(\delta_\A \left(1+ (\rho+\delta_\c) \what L_M(\delta_\A) \right) + (1+L) \delta_\c\right)^2\tr(\boldsymbol{\Sigma}_0)/N
		\end{aligned}
		\end{equation}
	}with probability at least $1-\varepsilon_\A-\varepsilon_\c$, where
	\[\what L_M(\delta)=\max_{1\le i \le M}\; \frac{(L+\delta)^i-L^i}{\delta}\]
	\begin{equation}
	\begin{aligned}
	\varepsilon_\A &\sim d_1 \exp\{-O(K)\}\\
	\varepsilon_\c &\sim 2\exp\{ -O(K^{1-2\nu})\}\\
	\delta_\A &=O\left(\sqrt{\log N/K}\right)\\
	\delta_\c &=O\left(\sqrt{\log N/K^{2(\nu-\beta)}}\right).
	\end{aligned}
	\end{equation}
\end{thm}
\begin{proof}[Proof of Theorem 1]
	First, applying the union bound to results from Lemmas~\ref{lem:CGP_estim1} and~\ref{lem:CGP_c_body}, we see that
	\[
	P\left( \left(\|\what\A-\A\|_2\le \delta_\A \right) \bigcap \left( \|\what\c-\c\|_1\le \delta_\c \right) \right)\ge 1- \varepsilon_\A-\varepsilon_\c.
	\]
	Thus, we proceed using $\|\what\A-\A\|_2\le \delta_\A$ and  $\|\what\c-\c\|_1\le \delta_\c$. Then, $\|\what\A\|_2\le \|\A+(\what\A-\A)\|_2\le L+\delta_\A$. Similarly, $\|\what\c\|_1\le \rho+\delta_\c$. Then,
	\begin{align}
	\max_{1\le i\le M}&\|\what\A^i-\A^i\|_2& \nonumber\\
	&\le \|\what\A-\A\|_2\max_{0\le i\le M\!-\!1} \left\|\sum\limits_{j=0}^{i} \what\A^j\A^{i-j}\right\|_2& \nonumber\\
	&\quad\le \delta_\A \max_{0\le i\le M\!-\!1}\sum\limits_{j=0}^{i}  \left\|\what\A\right\|_2^j\left\|\A\right\|_2^{i-j}&\\
	&\quad\le \delta_\A \max_{0\le i\le M\!-\!1}\sum\limits_{j=0}^{i}  (L+\delta_\A)^j L^{i-j}& \nonumber\\
	&\quad\le \delta_\A \what L_M(\delta_\A).& \nonumber
	\end{align}
	Note that $\what L_M(\delta)=O(ML^{M-1})$ when $\delta\rightarrow 0$.
	
	Next, dropping from the list of arguments for the function $f(\A,\c,\X'_{k-1})$ defined below equation~\eqref{eq:CGP_error_metric} the explicit dependence on $\X_{k-1}'$ for compactness of notation,
	{\small
	\begin{equation*}
	\begin{aligned}
	\Ebb \!& \left[ \| \x[k] \!-\! f(\what\A, \what\c) \|_2^2 \right] \!-\! \Ebb\left[\|\x[k]\!-\!f(\A,\c)\|_2^2\right]\\
	&= \Ebb \left[ \left( f(\A,\c) \!-\! f(\what\A, \what\c) \right)^\top \left( 2\x[k] \!-\! f(\A,\c) \!-\! f(\what\A, \what\c) \right) \right]\\
	&\overset{(a)}{=} \Ebb\left[\left(f(\A,\c)\!-\!f(\what\A,\what\c) \right)^\top\left(2\w[k]\!+\!f(\A,\c)\!-\!f(\what\A,\what\c)\right)\right]\\
	&=\Ebb\left[\|f(\what\A,\what\c) \!-\! f(\A,\c)\|_2^2 \right]\\
	&= \Ebb\left[\|f(\what\A,\what\c)\!-\!f(\A,\what\c) \!+\! f(\A, \what\c) \!-\! f(\A,\c)\|_2^2\right]\\
	&\le \Ebb \left[ \left(\|f(\what\A,\what\c) \!-\! f(\A,\what\c) \|_2 \!+\! \|f(\A,\what\c)\!-\!f(\A,\c)\|_2\right)^2\right]\\
	&\overset{(b)}{\le} \!\left(\|\D(\what\A,\what\c) \!-\! \D(\A,\what\c) \|_2 \!+\! \|\D(\A,\what\c) \!-\! \D(\A,\c)\|_2 \right)\!^2 \Ebb \left[ \|\X_{k\!-\!1}'\|_2^2\right]\\
	&=\Big(\underset{V_1}{\underbrace{\|\D(\what\A,\what\c) \!-\! \D(\A,\what\c) \|_2}} \!+\! \underset{V_2}{\underbrace{\|\D(\A,\what\c\!-\!\c)\|_2}} \Big)^2 M\tr(\boldsymbol{\Sigma}_0),
	\end{aligned}
	\end{equation*}
	}where $\D(\A',\c')=\left(\begin{array}{cccc}\A' & P_1(\A',\c') & \ldots & P_M(\A',\c') \end{array}\right)$, and where $(a)$ is due to $\x[k]-f(\A,\c)=\w[k] \perp \x[j] \;\forall j< k $, and $(b)$ is due to $\D(\A',\c')\X'_{k-1} \!=\! f(\A',\c')$.
	Now consider the term $V_1$,
	\begin{equation*}
	\begin{aligned}
	V_1 \!&\le\! \sum\limits_{i=1}^{M} \! \|P_i(\what\A,\what\c) \!-\! P_i(\A,\what\c)\|_2 \!\le\! \delta_\A \!+\! \sum\limits_{i=2}^{M} \! \sum\limits_{j=1}^i|\what c_{ij}|\|\what\A^j \!-\! \A^j\|_2 \\
	\!&\le \!\delta_\A + \delta_\A \what L_M(\delta_\A)\|\what\c\|_1  \le \delta_\A \left(1+ (\rho+\delta_\c) \what L_M(\delta_\A) \right).
	\end{aligned}
	\end{equation*}
	Next we similarly bound $V_2$,
	\begin{equation*}
	\begin{aligned}
	V_2 &\le \sum\limits_{i=1}^{M}\|P_i(\A,\what\c-\c)\|_2 \le \sum\limits_{i=2}^{M}\sum\limits_{j=0}^i |\what c_{ij}-c_{ij}|\|\A^j\|_2 \\
	&\le \sum\limits_{i=2}^{M} |\what c_{i0}-c_{i0}|+\sum\limits_{i=2}^{M}\sum\limits_{j=1}^i |\what c_{ij}-c_{ij}|\|\A^j\|_2 \\
	&\le \|\what\c-\c\|_1+L\|\what\c-\c\|_1 \overset{(a)}{\le} (1+L) \delta_\c.
	\end{aligned}
	\end{equation*}
	where $(a)$ follows from $\|\what\c-\c\|_1\le\delta_\c$ proven in lemma~3, and the result follows.
\end{proof}

\end{document}